\documentclass[11pt,a4paper]{article}

\usepackage{jheppub}
\usepackage{color}
\usepackage[T1]{fontenc} 
\usepackage{lmodern}
\usepackage{url}
\usepackage{xcolor}
\usepackage{subcaption}
\usepackage{float}
\usepackage{amsmath,amssymb}
\usepackage{amsfonts}
\usepackage{rotating}
\usepackage{latexsym}
\usepackage{bm}
\usepackage[utf8]{inputenc}
\usepackage{float}
\usepackage{ulem}
\usepackage{hyperref}
\usepackage{cite}
\usepackage{braket}
\usepackage[nopar]{lipsum}
\hypersetup{ setpagesize=false,  colorlinks=true, linkcolor=blue, citecolor=red, linktocpage=true, urlcolor=blue, hypertexnames=true}
\usepackage{booktabs}
\usepackage{cancel}
\usepackage{bbold}
\usepackage{cleveref}
\usepackage{changepage} 
\usepackage{multirow}
\usepackage[section]{placeins}

\newcommand{\snr}{{\rm SNR}}
\newcommand{\sr}{{\rm SR}}
\newcommand{\detun}{\Delta_\omega}

\newcommand{\paren}[1]{\left( #1 \right)}

\newcommand{\Fig}[1]{Fig.~\ref{#1}}
\newcommand{\Eq}[1]{Eq.~(\ref{#1})}
\newcommand{\Eqs}[2]{eqs.~(\ref{#1}) and (\ref{#2})}
\newcommand{\Sec}[1]{section~\ref{#1}}

\newcommand{\App}[1]{Appendix~\ref{#1}}

\newcommand{\DMphase}{\varphi}

\graphicspath{{figures/}}

\title{On the Speed-up of Wave-like Dark Matter Searches with Entangled Qubits}

\author[1,2]{Arushi Bodas,}
\author[1]{Sohitri Ghosh,}
\author[1]{Roni Harnik}
\affiliation[1]{Theoretical Physics Division, Fermi National Accelerator Laboratory, Batavia, 60510, IL, USA}
\affiliation[2]{Enrico Fermi Institute and Leinweber Institute for Theoretical Physics, University of Chicago, Chicago, IL 60637, USA}

\emailAdd{arushib@uchicago.edu}
\emailAdd{sohitri@fnal.gov}
\emailAdd{roni@fnal.gov}

\abstract{
Qubit-based sensing platforms offer promising new directions for wave-like dark matter searches.
Recent proposals demonstrate that entangled qubits can achieve quadratic scaling of the signal in the number of qubits. 
In this work we expand on these proposals to analyze the bandwidth and scan rate performance of entangled qubit protocols across different error regimes.
We find that the phase-based readout of entangled protocols preserves the search bandwidth independent of qubit number, in contrast to power-based detection schemes, thereby achieving a genuine scan-rate advantage. 
We derive coherence time and error rate requirements for qubit systems to realize this advantage.
Applying our analysis to dark photon searches, we find that entangled states of approximately 100 qubits can become competitive with benchmark photon-counting cavity experiments for masses $\gtrsim 30{-}40~\mu{\rm eV}$, provided sufficiently low error rates are achieved.
The advantage increases at higher masses where cavity volume scaling becomes less favorable.}

\begin{document}

\preprint{{\footnotesize FERMILAB-PUB-25-0719-T}}

\maketitle
\flushbottom

\section{Introduction}

Despite decades of evidence for the existence of dark matter (DM), its fundamental nature remains one of the greatest mysteries of physics. Wave-like dark matter candidates, particularly axions~\cite{Peccei:1977hh, Dine:1981rt, Preskill:1982cy} and dark photons~\cite{Holdom:1985ag, Nelson:2011sf}, are well-motivated extensions to the Standard Model that can address outstanding theoretical problems. These candidates couple to electromagnetic fields, making them accessible to laboratory searches through their interactions with photons and charged particles.
Specifically for light wave-like DM, which can be treated as a classically oscillating field at a frequency set by its unknown mass, the most widely used experimental strategies rely on resonant detection, such as the haloscope design~\cite{Sikivie:1983ip}, where dark matter converts to detectable photons in the presence of strong magnetic fields in electromagnetic cavities. Because resonant interactions are active only within narrow frequency ranges, these experiments must scan across broad frequency spaces to cover viable parameter regions. This scanning process is inherently time-consuming, making the scan rate at a given sensitivity depth an important metric for comparing experimental approaches.

Qubits represent an alternative class of resonant detectors to cavities, absorbing and emitting energy when driven at their transition \cite{Chen:2022quj,Nakazono:2025tak}. 
As opposed to linear resonators, qubits have nonlinear level spacing and can thus be quantum-controlled more effectively
\cite{Devoret2013,Krantz2019}. Modern qubit architectures span frequency ranges relevant 
for light dark matter searches, from superconducting circuits operating at GHz frequencies 
\cite{Koch:2007hay,Clarke2008,Blais2021} to trapped ions reaching optical scales 
\cite{BlattWineland2008,Bruzewicz2019}. Given the rapid development of quantum technologies 
driven by quantum computing applications, it is interesting 
to explore their potential for fundamental physics searches. Though individual 
qubits can act as powerful sensors, it is interesting to utilize entanglement among many 
qubits as a resource to improve such searches \cite{Giovannetti2004,Giovannetti2011,Pezze2018}.

In going from one to $N$ independent detectors, whether cavities, qubits or other sensors, the signal power typically scales linearly with $N$ while noise scales as $\sqrt{N}$, yielding signal-to-noise ratio improvements of $\sqrt{N}$. Coherent amplitude addition could in principle achieve quadratic signal scaling ($N^2$).
An example of this was shown in Refs.~\cite{chen2023quantum, Ito:2023zhp, Chen:2024aya} for qubits prepared in entangled GHZ states~\cite{GHZ1989}\footnote{For another example of super-radiant rates, $N^2$-scaling, see~\cite{Arvanitaki:2024taq, Galanis:2025amc}.}. Interestingly, if the noise scaling remains as $\sqrt{N}$, this potentially provides an SNR improvement to scale as $N^{3/2}$. We note, however, that the effects in terms of scan rate have not been demonstrated as of yet.

This paper provides a comprehensive analysis of this entangled qubit approach. We demonstrate that the search bandwidth remains independent of qubit number in the entangled protocol, confirming a genuine scaling advantage over power-based detection schemes in terms of scan rate. We parametrize realistic noise sources and their scaling properties, derive scan rate expressions for both entangled and unentangled protocols, and establish parameter regimes where entangled approaches provide advantages. Finally, we compare superconducting qubit systems operating in the GHz range with a benchmark for single cavity experiments~\footnote{We choose to compare to a single cavity searching for single photons deposited in an empty cavity~\cite{dixit2021searching}. Other cavity based schemes, e.g.~\cite{Agrawal:2023umy, Kuo:2024duq, Zheng:2025qgv} may be compared to this benchmark.}, determining how many entangled qubits would be required to achieve competitive performance.

It is worthwhile to note that an $N^2$ scaling of signal is not common. Naively, one may hope that electric fields $E$ from individual cavities can be added coherently such that the power (and number of photons) $\propto E^2$ grow quadratically with the number of cavities. However, this approach consistently encounters a fundamental limitation: the higher sensitivity at a certain frequency comes at the cost of bandwidth reduction by $1/N$~\cite{Lasenby:2019hfz}. Therefore, the additional advantage from coherent addition disappears while evaluating scan rate, which is a more relevant figure of merit for frequency-scanning experiments, yielding performance equivalent to independent cavity operation. This reflects a deeper constraint arising from energy conservation: the total power extractable from dark matter scales linearly with detector volume, and coherent schemes merely redistribute this fixed power budget across frequency space~\cite{Blinov:2024jiz}.

Circumventing this limitation to achieving $N^2$ scaling requires measurement schemes that do not rely on power extraction. We demonstrate that the entangled qubit protocol achieves precisely this: dark matter interaction leaves the average energy unchanged while encoding information in relative phases between entangled state components. These phases can be extracted through interferometric readout techniques, preserving both quadratic signal scaling and full bandwidth.

Quantum technologies remain in early stage of development, and creating high-fidelity entangled states across many qubits presents significant experimental challenges. Our aim is not to propose immediately realizable experiments, but rather to establish theoretical benchmarks and scaling laws that can guide future experimental efforts as quantum technologies mature. By understanding both the advantages and limitations of entangled detection protocols, we can better assess the potential of quantum sensors for dark matter discovery as the underlying technology develops.

The rest of the paper is organized as follows.
In \Sec{sec:entanglement_bases_sensing}, we review dark matter-qubit interactions and the entangled detection protocol. We also analyze its bandwidth scaling and demonstrate the absence of the typical bandwidth-sensitivity tradeoff. 
In \Sec{sec:noise_and_scanrate}, we characterize noise sources, derive scan rates for qubit-based protocols, and analyze their scaling behavior in different regimes. 
Section~\ref{sec:cavity_qubit_comparison} provides a comparison of entangled-state protocol with cavity-based experiments for dark photons in GHz frequency range. 
We conclude in \Sec{sec:conclusion} with a summary and future directions.

\section{Entanglement-Based Dark Matter Sensing}
\label{sec:entanglement_bases_sensing}

We review an entanglement-based dark matter (DM) sensing protocol that was proposed in refs.~\cite{chen2023quantum, Ito:2023zhp} starting with the interaction of DM with a single qubit, and continuing to present the protocol for $n_q$ entangled qubits.
We also analyze the bandwidth of the protocol, which is important for deriving the scan rate in \Sec{sec:noise_and_scanrate}.

\subsection{Dark matter-qubit interaction}
\label{sec:dm_interaction}

Light DM can be modeled as a classical, spatially coherent field oscillating at a frequency determined by its mass $m_{\rm DM}$.
This classical treatment is justified when the occupation number of DM particles is large, as expected for $m_{\rm DM} \ll 10$ eV. The field exhibits temporal coherence over a characteristic time scale $\tau_\mathrm{\small DM} \sim 1/(m_{\rm DM} v^2)$, where $v \sim 10^{-3}$ is the virial velocity in our galaxy. For $m_{\rm DM} \sim \mu$eV, the corresponding coherence time is $\sim$ millisecond.

Light DM candidates that couple to electromagnetic fields can source oscillating electric fields of the general form:
\begin{equation}
\vec{E}(\vec{x} , t) = \epsilon_{\rm DM} \sqrt{2\rho_\mathrm{DM}} \cos (m_{\rm DM}t - \varphi) \hat{n}(\vec{x}),
\end{equation}
where $\epsilon_{\rm DM}$ characterizes the coupling strength between DM and electromagnetic fields, $\rho_{\rm DM}$ is the local dark matter density, $\varphi$ is a time-dependent phase, and $\hat{n}(\vec{x})$ specifies the polarization direction of the induced electric field.

The finite coherence of the DM field manifests through the random phase $\varphi$, which changes by order unity every coherence time $\tau_\mathrm{\small DM}$. Since the virial velocity is small, the coherence length $\lambda_{\rm coh} \sim 1/(m_{\rm DM} v)$ is typically much larger than the experimental size.
For example, for $m_{\rm DM}\sim \mu{\rm eV}$, $\lambda_{\rm coh}\sim 200 \, {\rm m}$.
This allows us to neglect spatial gradients across the detector volume and treat all qubits as experiencing the same DM field, which is a crucial property for entanglement-based detection protocols.

We consider two well-motivated light DM candidates that couple to photons:\\
\textbf{Dark Photons:} In the simplest extensions of the Standard Model (SM), a dark photon ($A'$) is the gauge boson of a new $U(1)'$ symmetry that acquires mass through the St\"uckelberg mechanism. 
The dark photon can linearly mix with the SM photon through a gauge-kinetic mixing term:
\begin{equation}
\mathcal{H}_\mathrm{mix} = \epsilon_{A'} \left( \vec{E}\cdot \vec{E}' + \vec{B}\cdot \vec{B}' \right).
\end{equation}
For dark photons, the general electromagnetic coupling takes the specific form $\epsilon_{\rm DM} = \epsilon_{A'} $ (the gauge-kinetic mixing parameter) and $\hat{n}(\vec{x}) = \hat{n}_{A'}(\vec{x})$ (the dark photon polarization direction).\\
\textbf{Axions:} Axion-like particles (henceforth referred to as axions) can couple to photons nonlinearly through the interaction $g_{a\gamma} a F \tilde{F}$. 
In the presence of a static magnetic field $\vec{B}$, this coupling induces the electromagnetic field with $\hat{n}(\vec{x})$ in the magnetic field direction $ \hat{n}_{B}(\vec{x})$ and $\epsilon_{\rm DM} = \epsilon_a$ with
\begin{equation}
\epsilon_a = f(m_{\rm DM}R) g_{a \gamma} \frac{|\vec{B}|}{m_{\rm DM}}.
\end{equation}
The function $f(m_{\rm DM}R)$ accounts for the spatial profile of the magnetic field within the detection volume and depends on the ratio of the DM Compton wavelength to the characteristic size $R$ of the magnetic field region.

The oscillating DM-induced electric field acts as an external drive on the  dipole transition of the qubit. This interaction can be captured by an effective time-dependent Hamiltonian:
\begin{align}\label{eq:DM-qubit_Heff}
{\cal H}_{\rm eff} = \omega |1\rangle \langle 1|- 2 \eta \cos(m_{\rm DM} t-\varphi)\, (|1\rangle \langle 0|+|0\rangle \langle 1|),
\end{align}
where $|0\rangle$ and $|1\rangle$ are the ground and excited states of the qubit, $\omega$ is the qubit transition frequency, and $\eta$ is the effective Rabi frequency characterizing the interaction strength:
\begin{align}
\eta = \frac{1}{2} e \epsilon_{\rm DM}\, \kappa \sqrt{\rho_\mathrm{DM}V_\mathrm{eff} \omega} \cos \theta.
\end{align}
here $\theta$ is the angle between the dipole moment of the qubit and the electric field polarization, $V_\mathrm{eff}$ is the effective interaction volume, and $\kappa$ captures the electromagnetic mode overlap with the qubit (typically $\kappa \gtrsim {\cal O}(1)$).

The effective volume $V_\mathrm{eff}$ depends on the qubit implementation, but it is typically smaller than the corresponding electromagnetic cavity.
For superconducting qubits, $V_\mathrm{eff} \sim d^2 C$ where $d$ is the capacitor plate separation and $C$ is the capacitance. 
Typical values for a qubit at GHz frequency ($d \sim 100-300~\mu\mathrm{m}$, $C \sim 0.1~\mathrm{pF}$) gives $V_\mathrm{eff} \sim 10^{-4}-10^{-3}~\mathrm{cm}^3$, which is much smaller compared to an equivalent cavity with a volume of approximately $\sim m_{\rm DM}^{-3} \sim 10^3 ~ {\rm cm}^3$. 
For ion trap qubits, $V_{\rm eff} = (m_{\rm ion} \omega^2)^{-1}$, which is also smaller since $m_{\rm ion}\sim {\rm {\cal O}(10-100) \, GeV} \gg m_{\rm DM}$ for any light DM candidate.
However, qubits can achieve sensitivity to much smaller signal amplitudes
and can benefit from entanglement enhancement, as we will see below.

The DM interaction drives Rabi oscillations between the ground and excited states of the qubit. To analyze this evolution, we work in the rotating frame where the natural qubit evolution $e^{-i\omega t}|1\rangle$ is factored out. Writing the state as:
\begin{align}
| \psi(t) \rangle = \psi_{0}(t) | 0 \rangle +e^{-i\omega t} \psi_{1}(t) | 1 \rangle ,
\end{align}
the evolution equations become:
\begin{align}
\begin{pmatrix} \dot{\psi}_{0}(t)\\ \dot{\psi}_{1}(t) \end{pmatrix} &= i \begin{pmatrix} 0 & \eta  e^{-i(\Delta_\omega t+\varphi)}\\ \eta e^{i(\Delta_\omega t +\varphi)} & 0 \end{pmatrix} \begin{pmatrix} \psi_{0}(t)\\ \psi_{1}(t) \end{pmatrix},
\end{align}
where $\Delta_\omega \equiv \omega - m_{\rm DM}$ is the detuning from resonance. This derivation uses the rotating wave approximation, valid when $\Delta_\omega \ll m_{\rm DM}, \omega$, which neglects rapidly oscillating terms at frequency $\omega + m_{\rm DM}$.

The solution can be expressed as a unitary time evolution operator:
\begin{equation}
\begin{pmatrix} \psi_{0}(t)\\ \psi_{1}(t) \end{pmatrix} = U_\mathrm{DM} (t) \begin{pmatrix} \psi_{0}(0)\\ \psi_{1}(0) \end{pmatrix},
\end{equation}
where:
\begin{align}\label{eq:U_DM_main_text}
U_{\rm DM} &= \begin{pmatrix} e^{- it \Delta_\omega/2} (\cos \delta + i \sin \delta \sin \theta) & i e^{-i \varphi} e^{- it \Delta_\omega/2} \cos \theta \sin \delta \\ i e^{i \varphi} e^{it \Delta_\omega/2} \cos \theta \sin \delta & e^{ it \Delta_\omega/2} (\cos \delta - i \sin \delta \sin \theta) \end{pmatrix} \\ \notag
& {\rm with } \quad \delta = \frac{\sqrt{\Delta_\omega^2+ 4 \eta^2}}{2} t , \quad \theta = \tan^{-1} \left(\frac{\Delta_\omega}{2 \eta}\right).
\end{align}
On resonance ($\Delta_\omega=0$), this simplifies to:
\begin{align}\label{eq:U_DM_on_res} 
    U_{\rm DM} &= \begin{pmatrix}
       \cos \delta  & i e^{-i \varphi} \sin \delta \\
        i e^{i \varphi} \sin \delta & \cos \delta  
    \end{pmatrix} \qquad {\rm with} \quad \delta = \eta \, t\, .
\end{align}
For a qubit initially in the ground state, the interaction with the DM leads to a non-zero amplitude on the excited state after time $t_{\rm exp} (\lesssim \tau_{\rm MD})$,  providing the basic signal for DM detection.
The signal probability is 
\begin{align}
\label{eq:Psig}
    p_{\rm sig} = |\psi_1 (t_{\rm exp})|^2 = \sin^2\delta \approx (\eta \,t_{\rm exp})^2.
\end{align}

In the off-resonance limit $\Delta_{\omega} \gg \eta$, using \Eq{eq:U_DM_main_text}, we get  $p_{\rm sig} \sim (2 \eta^2/\Delta_{\omega}^2)\sin(\Delta_{\omega} t_{\rm exp}/2)$, assuming 
$t_{\rm exp} \lesssim \tau_\mathrm{\small DM}$ such that $\varphi$ remains constant throughout the experiment. We can see that for a single qubit, the detuning beyond which the signal drops considerably, or the bandwidth, is $\sim 1/t_{\rm exp}$.

\subsection{GHZ entangled-state protocol}
\label{sec:entangled_protocol}

\begin{figure}
    \centering
    \includegraphics[width=0.95\linewidth]{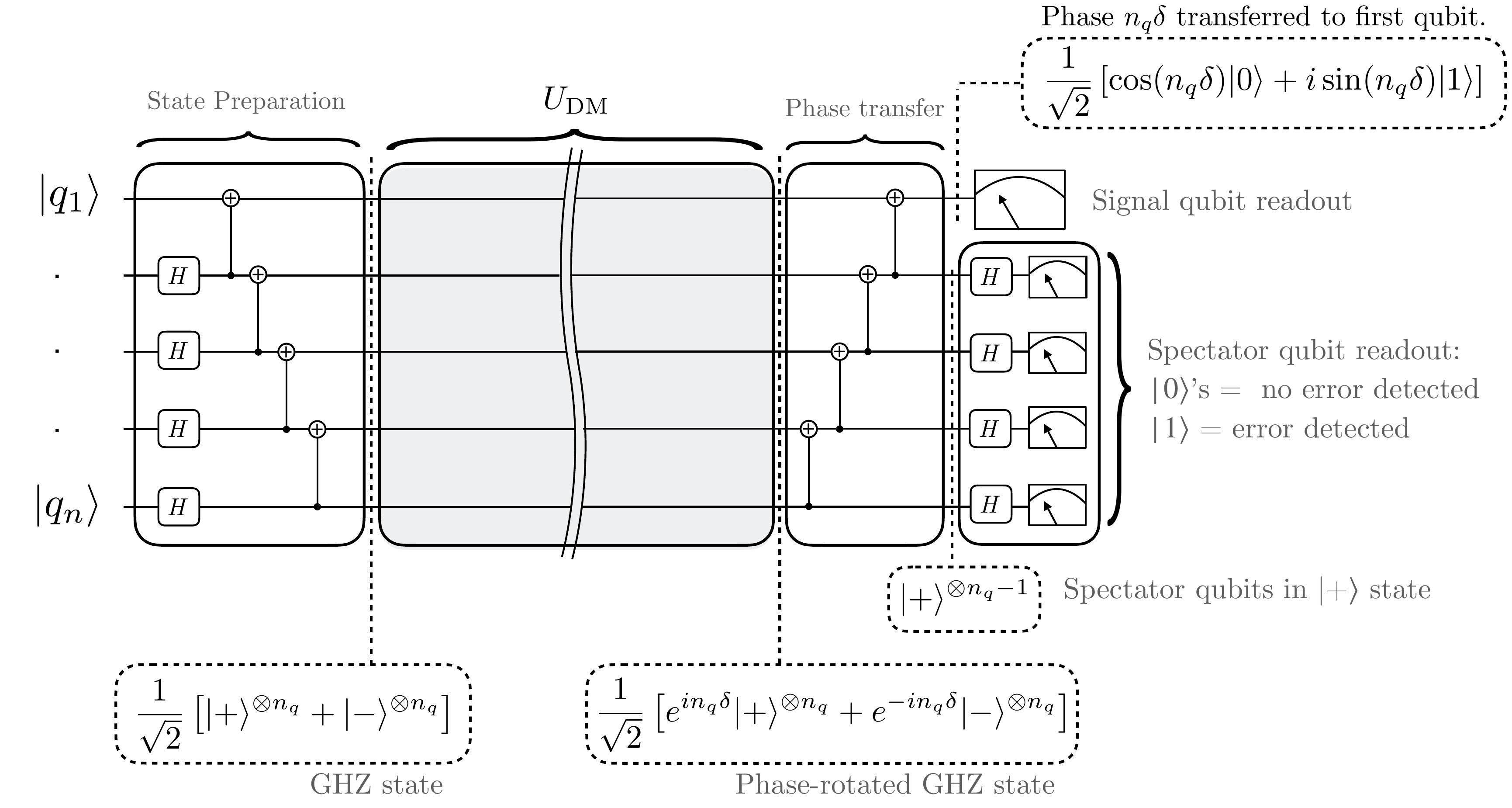}
    \caption{Quantum circuit implementing the entangled-state protocol for DM detection. The circuit shows: (i) preparation of GHZ-like entangled state using Hadamard and CNOT gates, (ii) exposure to DM (the $U_\mathrm{DM}$ ``gate''), and (iii) phase transfer via CNOT gates to transfer the phase information to the signal qubit for measurement. In the absence of DM the signal qubit will be in $|0\rangle$. The other (``spectator'') qubits will be in $|0\rangle$ states at the end the protocol, but can be monitored to detect errors.
    Dotted boxes show the state of the system in various stages of the protocol.
    \label{fig:DM-circuit}}
\end{figure}

The single-qubit analysis reveals a key insight that motivates an entangled detection strategy. On resonance ($\Delta_\omega = 0$) and with $\varphi = 0$, the dark matter evolution operator from \Eq{eq:U_DM_on_res} has eigenstates $|\pm\rangle = \frac{1}{\sqrt{2}}(|0\rangle \pm |1\rangle)$ with eigenvalues $e^{\pm i \delta}$, where $\delta = \eta t$. Dark matter therefore adds a phase $\pm \delta$ to these eigenstates while leaving their populations unchanged.
This phase-only interaction suggests an interferometric detection strategy: if an entangled superposition of many qubits in these eigenstate configurations is prepared, each qubit will accumulate the same phase $\delta$, leading to a total phase enhancement of $n_q \delta$ that can be read out interferometrically.

A protocol achieving exactly this coherent phase accumulation was proposed in refs.~\cite{chen2023quantum, Ito:2023zhp} which focused on superconducting qubits and trapped ions respectively. The quantum circuit schematic is shown in \Fig{fig:DM-circuit} and consists of three main stages:\\[6pt]
\textbf{GHZ state preparation:}
Starting with $n_q$ qubits in their ground states, a GHZ-like entangled state is prepared:
\begin{equation}\label{eq: GHZ_state}
    \ket{0}^{\otimes n_q} \xrightarrow{\rm state ~ preparation} \frac{1}{\sqrt{2}} \left(\ket{+}^{\otimes n_q}+\ket{-}^{\otimes n_q}\right).
\end{equation}
This superposition places all qubits in the eigenbasis of the DM interaction, with each component ($\ket{+}^{\otimes n_q}$ and $\ket{-}^{\otimes n_q}$) ready to accumulate opposite phases under DM evolution. The state preparation can be implemented using Hadamard gates followed by nearest-neighbor CNOT gates (as shown in the figure). However, this protocol is by no means unique~\cite{Monz2011, Song2019,bao2024creating}.\\[6pt]
\textbf{DM interaction:}
The entangled state is left to evolve under the influence of DM for time $t_{\rm exp}$. Due to the large coherence length of DM, all qubits experience the same DM field and undergo identical evolution under $U_{\rm DM}$. 
The phases from all $n_q$ qubits add coherently within each component of the superposition:
\begin{equation}\label{eq: postDM_state}
    \frac{1}{\sqrt{2}} \left(\ket{+}^{\otimes n_q}+\ket{-}^{\otimes n_q}\right)\xrightarrow{U_{\rm DM}^{\otimes n_q}} \frac{1}{\sqrt{2}} \left(e^{i n_q \delta}\ket{+}^{\otimes n_q}+e^{-i n_q \delta} \ket{-}^{\otimes n_q}\right).
\end{equation}
\textbf{Phase transfer and measurement:}
Direct measurement in the computational basis at this stage would require constructing a complicated observable that preserves $n_q$-enhanced phase information.
Instead, a series of CNOT gates can be employed to transfer all phase information to the first qubit while leaving the remaining $(n_q-1)$ qubits in product states. This ``phase transfer'' operation maps the state to the following:
\begin{equation}
\begin{split}
    \frac{1}{\sqrt{2}} \left(e^{i n_q \delta}\ket{+}^{\otimes n_q}+e^{-i n_q \delta} \ket{-}^{\otimes n_q}\right) &\xrightarrow{\rm phase ~ transfer} \frac{1}{\sqrt{2}} \left(e^{i n_q \delta}\ket{+}+ e^{-i n_q \delta}\ket{-}\right) \ket{+}^{\otimes n_q -1} \\
    &=  \frac{1}{\sqrt{2}} \left( \cos{(n_q \delta)} \ket{0}+ i \sin{(n_q \delta)}\ket{1} \right) \otimes \ket{+}^{\otimes n_q -1}.
    \end{split}
\end{equation}
The series of Hadamard gates on the spectator qubits at the end turn turn them all into $|0\rangle$. 
The probability of detecting the first qubit in the excited state provides our DM signal. In the weak signal limit,
\begin{equation}
\label{eq:Pexcite_on_resonance_alpha0}
    p_{\rm sig}^{\rm ent} = \sin^2 (n_q \delta) \approx n_q^2 \delta^2.
\end{equation}
Comparing this to $n_q$ independent qubits, where each has signal probability $p_{\rm sig}^{\rm single} = \delta^2$ and the total signal scales as $n_q \delta^2$, we see that the entangled protocol provides a quadratic enhancement.
The more general case of $\DMphase \neq 0$ only adds a factor of $\cos^2\varphi$ as shown in \Eq{eq:Pexcite_onResonance}, and retains the $n_q^2$ scaling. 

\subsection{Bandwidth of the entangled-state protocol}
\label{sec:bandwidth}

As explained before, scan rate is an important criterion for any resonance-based search, which includes bandwidth of each measurement.
Multi-component detection systems typically exhibit bandwidth that depends on the number of subsystems. Examples include dielectric stacks (MADMAX~\cite{Millar:2016cjp}, LAMPOST~\cite{Baryakhtar:2018doz}), photonic systems~\cite{Blinov:2024jiz}, and plasma haloscopes (ALPHA~\cite{ALPHA:2022rxj}). In these power-based schemes, quadratic peak signal enhancement comes at the cost of bandwidth reduction by $1/N$, resulting in linear scaling of the frequency-integrated rate~\cite{Lasenby:2019hfz}. 
This is because the total power drawn from DM can only scale linearly with volume (see \App{app:volume_scaling} for derivation).
The entangled qubit protocol, by contrast, draws no average power from the DM field and stores information purely in a quantum phase. 
This can be seen from \Eq{eq: postDM_state} where the expectation value of the energy of the state before and after DM exposure is the same.
This fundamental difference in the two readout schemes suggests the
search bandwidth can potentially remain independent of $n_q$ in the GHZ setup, which we now demonstrate.

\begin{figure}[t]
    \centering
    \includegraphics[width=0.48
    \linewidth, height = 0.3\linewidth]{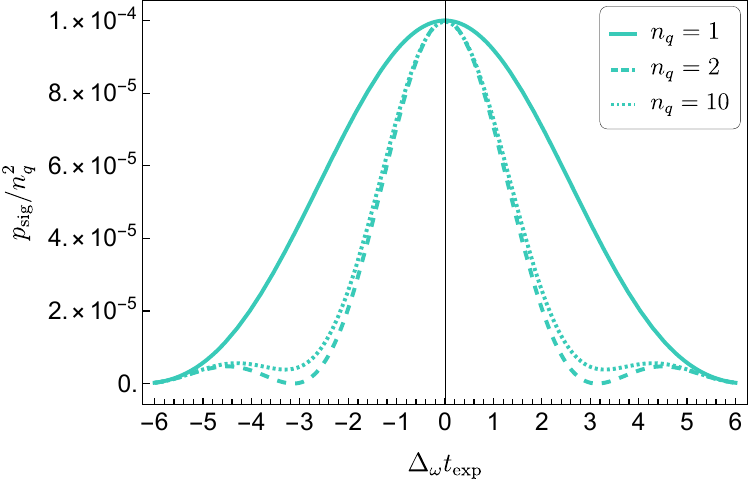}
    \hspace{0.3 em}
    \includegraphics[width=0.48\linewidth, height = 0.3\linewidth]{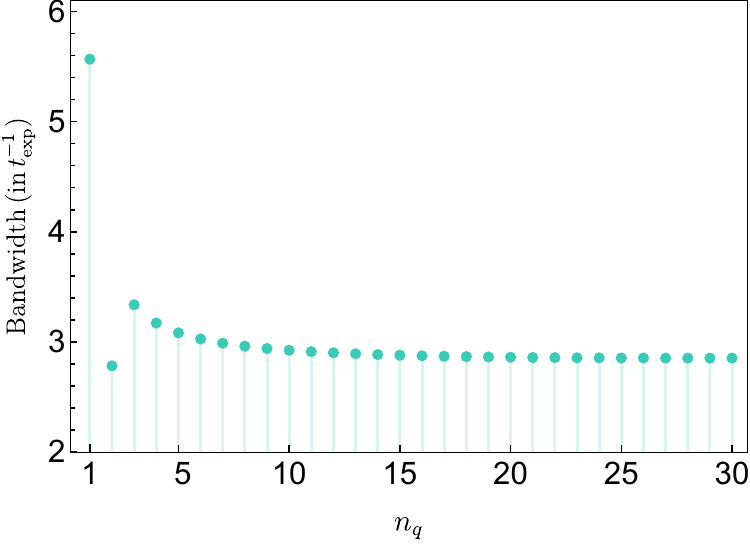}  
    \caption{Left: signal probability in the entangled-state protocol (normalized by $n_q^2$) as a function of detuning $\Delta_{\omega}$. Right: bandwidth of the entangled-state protocol as a function of $n_q$. We have fixed $\eta t_{\rm exp}= 0.01$ in both plots.}
    \label{fig:bandwidth_nq}
\end{figure}

\begin{figure}[hbt]
    \centering
    \includegraphics[width=0.6\linewidth]{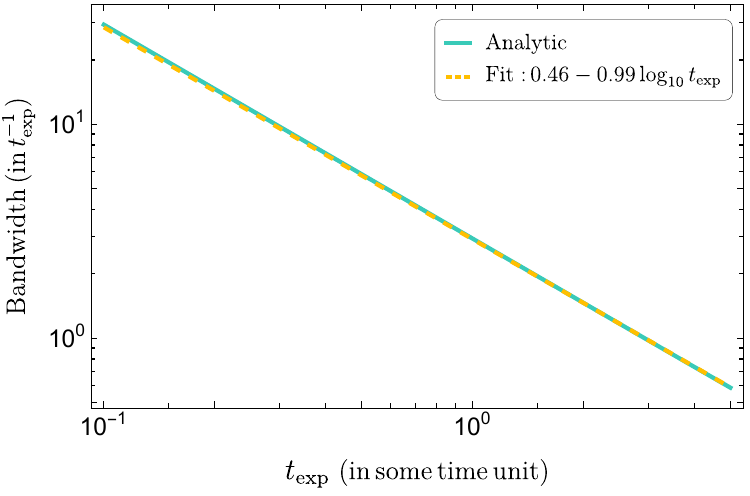}
    \caption{Bandwidth of the entangled-state protocol as a function of $t_{\rm exp}$ (solid cyan) and numerical fit (yellow dashed). We have fixed $n_q =10$ and assumed $t_{\rm exp} \leq \tau_\mathrm{\small DM}$.}
    \label{fig:bandwidth_vs_tDM}
\end{figure}

The bandwidth of the protocol can be determined by analyzing the off-resonance behavior. For detuning $\Delta_\omega \gg \eta$, the signal probability (derived in \App{app:excitation_prob}) takes the form:
\begin{align}
     p_{\rm sig} \xrightarrow{\rm off-res} n_q^2 \, \frac{4 \eta^2}{\detun^2} \sin^{2} \paren{\frac{\detun t_{\rm exp}}{2}}  \cos^{2}\paren{\DMphase+ \frac{\detun t_{\rm exp}}{2}}.
\end{align}
The key observation is that the signal drops when $\Delta_\omega \gtrsim t^{-1}_{\rm exp}$, indicating that the bandwidth scales as $\propto t_{\rm exp}^{-1}$. 
Crucially, the bandwidth is independent of $n_q$.
\Fig{fig:bandwidth_nq} demonstrates this result explicitly. The left panel shows the signal probability (normalized by $n_{q}^2$) as a function of the detuning for  different values of $n_q$ at fixed $t_{\rm exp}$ and $\varphi =0$.
The right panel shows the bandwidth (full width at half maximum) of the $p_{\rm sig}$-vs-$\Delta_\omega$ plot as a function of $n_q$.
Both of these confirm that the bandwidth has no dependence on $n_q$.
\Fig{fig:bandwidth_vs_tDM} further confirms the $ t_{\rm exp}^{-1}$ scaling by showing bandwidth as a function of exposure time for fixed $n_q = 10$, assuming $t_{\rm exp}\lesssim \tau_\mathrm{\small DM}$. The numerical fit (dashed line) validates the predicted scaling.

For $t >  \tau_\mathrm{\small DM}$, the bandwidth saturates to $1/\tau_\mathrm{\small DM}$. However, the amplitude gain is no longer linear in time due to the change in the DM phase $\varphi$ every coherence time, but it instead grows as $\propto \sqrt{(t/\tau_\mathrm{\small DM})}$.
Therefore, running a single experiment for longer than $\tau_\mathrm{\small DM}$ is equivalent to performing a series of experiments of $\tau_\mathrm{\small DM}$ duration each (and negligible reset time).
For a single qubit, we will therefore set $t_{\rm exp} = {\rm min}\{\tau_\mathrm{\small DM}, \tau_{q}\}$, where $\tau_q$ is the qubit coherence time. 
For the entangled-state protocol, the coherence time of the GHZ state $\tau_\mathrm{\small GHZ} < \tau_q$, and we take $t_{\rm exp} = {\rm min}\{\tau_\mathrm{\small DM}, \tau_\mathrm{\small GHZ}\}$. We note that the coherence time of the GHZ state typically scales as $\tau_q/n_q$.

\section{Noise and Scan Rates}
\label{sec:noise_and_scanrate}

Having understood the entangled-state protocol and its bandwidth behavior, we now analyze the scan-rate advantage in the presence of experimental errors. We also compare entangled-qubit setups with established cavity experiments in \Sec{sec:cavity_qubit_comparison} to assess when qubit protocols become competitive.

To evaluate scan rates, we adopt the following procedure. We fix a target sensitivity $\eta_*$ that is weak enough to require multiple experimental repetitions. The number of repetitions needed to reach this sensitivity with a desired signal-to-noise ratio (SNR) determines the total time $\Delta t$ spent scanning one frequency bin of width $\Delta \omega \sim 1/t_{\rm exp}$ (from \Sec{sec:bandwidth}). The scan rate is then:
\begin{equation}
    {\rm SR} \equiv \left.\frac{\Delta \omega}{\Delta t}\right|_{\rm \eta_*}.
\end{equation}

\subsection{Noise sources and error models}
\label{sec:noise}

Quantum devices inevitably encounter errors during operation, which can lead to false positive for the excitation on the measurement qubit.
For our analysis, it is sufficient to categorize them into three types: 
(1) \textit{readout errors} (with probability~$p_{\rm ro}$) from incorrect measurement outcomes, 
(2) \textit{thermal-like errors} (with probability~$p_{\rm th}$) from amplitude damping, dephasing, and Pauli errors that accumulate over time at rate $\tau_q^{-1}$, and 
(3) \textit{gate errors} (with probability~$p_{\rm g}$) from imperfect gate operations, dominated by two-qubit CNOT gates in the entangled protocol.
For a single qubit operating over time $t_{\rm exp} \approx \tau_\mathrm{\small DM}$ (assuming $\tau_q > \tau_\mathrm{\small DM}$), the thermal error probability is $p_{\rm th} = 1-e^{-\tau_\mathrm{\small DM}/\tau_q} \approx \tau_\mathrm{\small DM}/\tau_q$, and the total error probability is $p_{\rm error} = p_{\rm ro}+p_{\rm th}$. 
Most of these errors lead to a false positive with ${\cal O}(1)$ probability. 

In unentangled qubits, errors affect qubits independently. 
However, entangled protocols exhibit error propagation: any error on any of the $n_q$ qubits can corrupt the final measurement on the readout qubit. 
The effective error probability therefore increases as $n_q \, p_{\rm th}$ for small error rates. 
While this increases the false positive rate, the enhanced $n_q^2$ signal rate can still yield net sensitivity improvements, as we analyze below.

For well-calibrated systems, the average false positive rate acts as a known background. 
The noise limiting DM sensitivity comes from the \textit{variance} of this background rate, which sets the achievable signal-to-noise ratio (SNR) at each frequency.

\subsection{Unentangled qubits}

Let us first establish the baseline SNR and scan rate for unentangled qubits.
To simplify the discussion, we take the time of a single measurement to be approximately the exposure time of qubits to DM.\footnote{ 
In superconducting qubits,
typical gate times are $\sim  10-100$ ns, while the corresponding DM coherence time for masses $\sim 10^{-6}-10^{-5}$ eV is much larger $\sim 100-1000 \, \mu$s.
The same is not true for certain higher frequency qubits in trapped ion/ cold atom architectures. However, if one restricts to lower frequency collective modes, as done in \cite{Ito:2023zhp}, the gate operation time can still be shorter.}
With $n_{\rm rep}$ repetitions of a single qubit experiment, the total signal is $S \approx (1-p_{\rm error})\,p_{\rm sig} \,  n_{\rm rep} $.\footnote{We assume that the DM interaction is weak and error probabilities are small, such that the erasure of excitations by another error or DM interaction can be ignored.}
The efficiency factor $(1-p_{\rm error})$ accounts for the fact some errors completely erase information about DM from the final measurement (see \App{app:error} for such effects of thermal errors).
The noise for the measurement is $N \approx \sqrt{p_{\rm error} n_{\rm rep}}$, giving a signal-to-noise ratio of
\begin{align}
    \left. \frac{S}{N}\right|_{1} \approx \frac{(1-p_{\rm error})\, \eta^2 t_{\rm exp}^{2} n_{\rm rep}^{1/2}}{ \sqrt{p_{\rm error}}}.
\end{align} 
Following the procedure outlined at the beginning of the section, the scan rate (SR) at the target sensitivity $\eta_*$ with a desired SNR for a single qubit is 
\begin{equation}\label{eq:scanrate_1}
    {\rm SR}_{1} \equiv \left. \frac{\Delta \omega}{\Delta t} \right|_{1,\eta_{*}} \approx \frac{1/t_{\rm exp}}{n_{\rm rep} t_{\rm exp}} \rightarrow \frac{\eta_{*}^{4} t_{\rm exp}^2}{(\snr)^{2} p_{\rm error}}.
\end{equation}
For unentangled qubits, $t_{\rm exp} = {\rm min}\{ \tau_\mathrm{\small DM}, \tau_q \}$.
In the following, we will generally work in the regime where $\tau_{q} > \tau_\mathrm{\small DM}$, such that $t_{\rm exp} \approx \tau_\mathrm{\small DM}$.
However, it is straightforward to adapt to the case where the experimental time is limited by qubit coherence instead.

Since the entangled-state protocol requires $n_q$ qubits for a single measurement, a fair comparison with unentangled qubits would be to run $n_q$ qubits independently in \textit{parallel} (denoted by the superscript/subscript `un'). This reduces the amount of time spent on  repeating single qubit measurements, such that $\Delta t_{\rm un} = (n_{\rm rep} /n_q)\Delta t_{1} $, giving
\begin{align}\label{eq:scanrate_un_smallerror}
    {\rm SR}_{\rm un} \equiv \left. \frac{\Delta \omega}{\Delta t} \right|_{{\rm un},\eta_{*}} \xrightarrow{t_{\rm exp} \approx \tau_\mathrm{\small DM}} \frac{n_q \, \eta_{*}^{4} \tau_\mathrm{\small DM}^2}{(\snr)^{2} p_{\rm error}}.
\end{align}

\subsection{Entangled-qubit protocol} 

We now analyze SNR and the scan rate for the entangled protocol, where errors can more significantly affect the scaling behavior.
As discussed before, any error on any of the $n_q$ qubits can corrupt the final readout.
For example, the probability of at least one thermal-like error in the entangled-state protocol goes as 
$p_{\rm th}^{\rm ent} = 1-\paren{e^{-t_{\rm exp}/\tau_q}}^{n_q} \approx n_q\,  p_{\rm th}$.
The effective coherence time of the state $\tau_\mathrm{\small GHZ} = \tau_{q}/n_q$ is much shorter as expected. 
The total error probability is 
$p_{\rm error}^{\rm ent} \approx p_{\rm ro}+ (p_{\rm g} + p_{\rm th})n_q$, which must remain small for sufficient signal efficiency.

From \Eq{eq:Pexcite_on_resonance_alpha0}, the signal after $n_{\rm rep}^{\rm ent}$ repetitions is 
$S \approx (1-p_{\rm error}^{\rm ent}) \, n_{q}^{2} \eta^2 t_{\rm exp}^{2} \, n_{\rm rep}^{\rm ent}$.
The noise, on the other hand, goes as 
$N \approx \sqrt{p_{\rm error}^{\rm ent} n_{\rm rep}^{\rm ent}}$, 
giving a signal-to-noise ratio of
\begin{align}\label{eq:SNR_entangled}
    \left. \frac{S}{N}\right|_{\rm ent} \approx \frac{(1-p_{\rm error}^{\rm ent}) \, n_{q}^2 \eta^2 t_{\rm exp}^{2} \, (n_{\rm rep}^{\rm ent})^{1/2}}{\sqrt{p_{\rm error}^{\rm ent}}}.
\end{align}
In \Sec{sec:bandwidth}, we showed that the bandwidth for the entangled protocol is independent of $n_q$, and is still given by $\Delta \omega \sim 1/t_{\rm exp}$. 
Then the scan rate of the entangled protocol at sensitivity $\eta_*$ with the same SNR as before is
\begin{align}
\label{eq: scanrate_entangled_general}
    {\rm SR}_{\rm ent} \equiv\left. \frac{\Delta \omega}{\Delta t} \right|_{{\rm ent},\eta_{*}} \approx \frac{1/t_{\rm exp}}{n_{\rm rep}^{\rm ent} t_{\rm exp}} \approx \frac{(1-p_{\rm error}^{\rm ent})^2 \, n_q^4 \eta_{*}^{4} t_{\rm exp}^2}{(\snr)^{2} \, p_{\rm error}^{\rm ent}}.
\end{align}

The entangled protocol exhibits different scaling regimes depending on which errors dominate and how $t_{\rm exp}$ is limited.
For small $n_q$, the GHZ state coherence time $\tau_\mathrm{\small GHZ}$ exceeds $\tau_\mathrm{\small DM}$, allowing $t_{\rm exp} \approx \tau_\mathrm{\small DM}$.
When readout errors dominate over gate and thermal-like errors, i.e. for $n_q < p_{\rm ro}/(p_{\rm g} + p_{\rm th})$,  $p_{\rm error}^{\rm ent} \approx p_{\rm ro}$, and the scan rate scaling compared to the single qubit is
\begin{align}
   \frac{\sr_{\rm ent}}{\sr_{\rm 1}} \, \xrightarrow[n_q < p_{\rm ro} /(p_{\rm g}+p_{\rm th})]{\tau_\mathrm{\small GHZ} > \tau_\mathrm{\small DM}} \, n_q^4 .
\end{align}
Compared to the parallel unentangled protocol (\Eq{eq:scanrate_un_smallerror}), there is an advantage of $\sr_{\rm ent} /\sr_{\rm un} \propto n_q^3$.
Once the gate and thermal-like errors become dominant, the errors start scaling with $n_q$, giving
\begin{align}\label{eq:entangled_n3_scaling}
     \frac{\sr_{\rm ent}}{\sr_{1}} \xrightarrow[n_q > p_{\rm ro} /(p_{\rm g}+p_{\rm th})]{\tau_\mathrm{\small GHZ} > \tau_\mathrm{\small DM}} n_q^3 \paren{\frac{p_{\rm ro}+p_{\rm th}}{p_{\rm g}+p_{\rm th}}}.
\end{align}
Compared to the unentangled protocol, $\sr_{\rm ent} /\sr_{\rm un} \propto n_q^2$.

If thermal-like errors dominate, increasing $n_q$ will eventually leads to the entangled state coherence time to fall below $\tau_\mathrm{\small DM}$,
forcing us to shorten the exposure time to $t_{\rm exp} \approx \tau_\mathrm{\small GHZ}$ to keep $p_{\rm th}^{\rm ent}$ manageable.
Setting $t_{\rm exp} = \tau_\mathrm{\small GHZ} \log(1-p_{\rm max})^{-1}$ to cap the total error at $p_{\rm max}$:
\begin{equation}
    \frac{\sr_{\rm ent}}{\sr_{1}} \xrightarrow[n_q <  p_{\rm max}/ p_{\rm g}]{\tau_\mathrm{\small GHZ} < \tau_\mathrm{\small DM}} n_q^2 \paren{\frac{\tau_q}{\tau_\mathrm{\small DM}}}^2 \paren{\frac{p_{\rm ro}+p_{\rm th}}{p_{\rm max}}}.
\end{equation}
Note that $p_{\rm ro}$ and $p_{\rm th}$ are typically smaller than $p_{\rm max}$ to which $p_{\rm th}^{\rm ent}$ is fixed.
Compared to the parallel protocol for unentangled qubits, $\sr_{\rm ent} /\sr_{\rm un} \propto n_q$.

As $n_q$ is increased further, gate errors increase to eventually match $p_{\rm max}$.
This is the maximum qubit number, $n_q^{\rm max} =  p_{\rm max}/p_{\rm g}$, beyond which the entangled protocol does not provide an advantage as the efficiency becomes too small.
Beyond this point, the strategy would be to divide qubits into sets of $n_q^{\rm max}$ entangled protocols running in parallel. The scan rate beyond $n_q^{\rm max}$ would scale $\propto n_q$, similarly to the unentangled parallel protocol.
If the gate errors are dominant, $n_q^{\rm max} $ can be reached reached even before the $\tau_\mathrm{\small GHZ}$ from thermal-like errors becomes comparable to $\tau_\mathrm{\small DM}$. 
In this case, there is no intermediate regime with $n_q^2$ scaling.

\begin{figure}[t]
    \centering
\includegraphics[width=0.75\linewidth,trim={5cm 4.cm 5cm 4.5cm},clip]{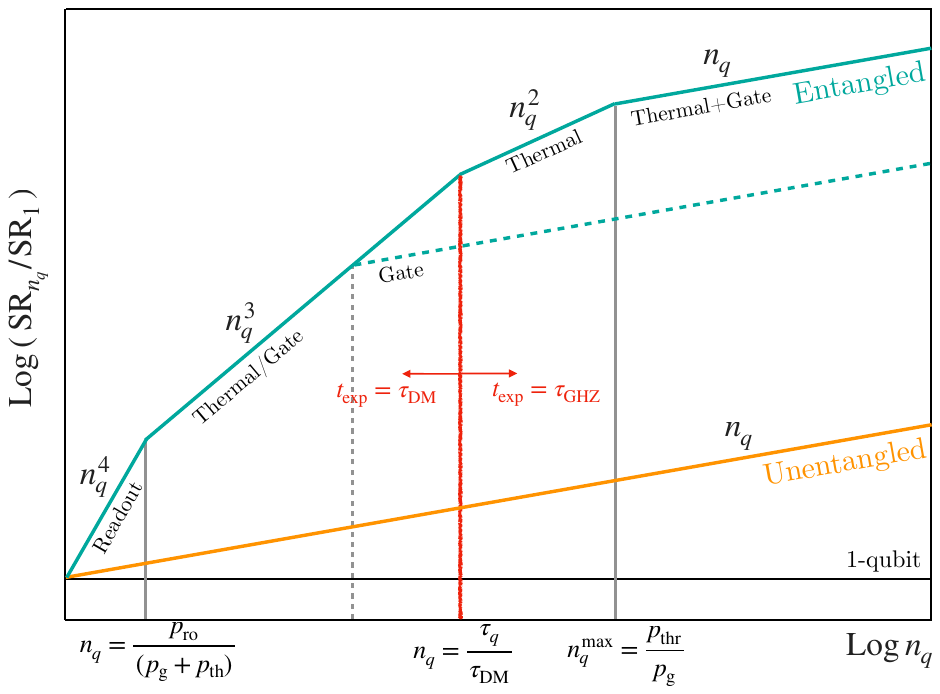}
    \caption{
    Scan rate scaling with $n_q$ for entangled (cyan) and unentangled parallel (orange) protocols. The entangled protocol exhibits following regimes: $n_q^4$ scaling when readout-error dominates, $n_q^3$ scaling when errors scale with $n_q$, $n_q^2$ scaling beyond the red line when $t_{\rm exp} \approx \tau_\mathrm{\small GHZ} < \tau_\mathrm{\small DM}$, and linear scaling beyond $n_q^{\rm max} = p_{\rm max}/p_{\rm g}$. Dashed lines show earlier transition for higher gate error rates.}
    \label{fig:scanrate_ratio}
\end{figure}

The various regimes discussed above are illustrated in \Fig{fig:scanrate_ratio} where the scan rates in the unentangled (parallel) and entangled protocols are taken in units of a single-qubit scan rate. 
We see that if error rates are sufficiently low, and qubit coherence times are large, the entangled protocol can provide significant advantage. Assuming small readout errors, the gain to the left of the red vertical line is $\sim \left(\tau_q/\tau_\mathrm{\small DM}\right)^3$. Namely, for qubit quality factors $Q_q$ that exceed DM quality factor $Q_{\rm DM} \sim 10^6$, say, by an order of magnitude, entanglement can provide a factor 1000 speedup over  single qubit (or a factor of 100 compared to the unentangled parallel sensing scheme).  Additional, more modest, gain is possible to the right of the red line by optimizing the per-shot exposure time.

We note that as the DM coherence time varies with DM mass, it is possible that even the single-qubit coherence time can fall below $\tau_\mathrm{\small DM}$ in the low mass regime, limiting the single qubit experiment time to $t_{\rm exp} \sim \tau_q$.
In this case, we are always on the left of the red line showing the boundary $\tau_\mathrm{\small GHZ} = \tau_\mathrm{\small DM}$ in \Fig{fig:scanrate_ratio}.
This will be relevant for \Fig{fig:scanrate_comparison}(b) in \Sec{sec:cavity_qubit_comparison}.

\subsection{Error characterization and mitigation}\label{sec:error}

The structure of the entangled protocol provides opportunities beyond simple error accounting.
The presence of spectator qubits, the $(n_q-1)$ qubits that are not directly measured, allows us to identify certain error types and measure error rates independently from the dark matter signal.
This capability is useful for characterizing systematic backgrounds in a dark matter search.
Table~\ref{tab:errors} summarizes how different error types affect both the signal qubit and spectator qubits.

\begin{table}[t]
    \centering
    \begin{tabular}{ccc}
    \hline\hline
       Error  & Signal qubit  & Spectator qubits \\
       \hline\hline
        no error & 100\% $|0 \rangle$ & all $|0\rangle$ \\
        \hline
        bit flip in $|q_j\rangle$ &  100\% $|1\rangle$  & all $|0\rangle$ \\ 
        \hline
        phase flip in $|q_j\rangle$ &  100\% $|0\rangle$  & all $|0\rangle$ \\
        \hline
        \\[-1.1em] 
        \multirow{4}{*}{decay of $|q_j\rangle$} &  50\% $|0\rangle $ &  
        $|q_j\rangle\otimes |q_{j+1}\rangle = \left\{
         \begin{array}{l}
             50\% |00\rangle   \\ 
             50\% |11\rangle  \end{array}\right.$ \\[1em] 
           &  50\% $|1\rangle $ & $|q_j\rangle\otimes |q_{j+1}\rangle = \left\{ 
           \begin{array}{l}
            50\% |00\rangle   \\ 50\% |11\rangle
        \end{array} \right.$ \\[1em]
        \hline
        \\[-1.1em] 
        \multirow{4}{*}{heating of $|q_j\rangle$} &  50\% $|0\rangle $ &  
        $|q_j\rangle\otimes |q_{j+1}\rangle = \left\{
        \begin{array}{l}
            50\% |00\rangle   \\ 50\% |11\rangle
        \end{array} \right.$ \\[1em] 
           &  50\% $|1\rangle $ & $|q_j\rangle\otimes |q_{j+1}\rangle = \left\{
           \begin{array}{l}
            50\% |00\rangle   \\ 50\% |11\rangle
        \end{array} \right.$ \\[1em]
        \hline\hline
    \end{tabular}
    \caption{Possible qubit errors and their effects on the signal and 
    spectator qubits in the entanglement-based protocol, assuming no DM signal.}
    \label{tab:errors}
\end{table}

Coherent errors act as unintended unitary operations without entangling the qubit with an external environment.
The two primary examples are bit flips (equivalent to an $X$ gate) and phase flips (equivalent to a $Z$ gate).
A bit flip error on any qubit in the system produces a false positive on the signal qubit without affecting the spectator qubits.
This is because the dark matter signal itself acts like a small-amplitude $X$ gate, so bit flip errors mimic the signal perfectly.
The indistinguishability of bit flip errors from true signals makes them particularly problematic, suggesting that qubit platforms with lower bit flip rates are preferred for this protocol.
Phase flip errors, by contrast, do not produce false positives.
As shown in ref.~\cite{chen2023quantum}, the entangled protocol naturally acts as an error-correcting code for phase flip errors.
This built-in protection against phase flips is a significant advantage of the entangled approach.

Incoherent errors involve the qubit entangling with an external environment, such as spontaneous decay (photon emission) or thermal excitation (photon absorption).
As derived in the \App{app:error}, when such an error occurs on qubit $i$ during the protocol, the signal qubit has 50$\%$ probability of showing a false positive, while spectator qubits $i$
and $i+1$ have 50$\%$ probability of both being excited. 
The outcome on the spectator measurements is not fully correlated with the outcome on the signal qubit, so they cannot be used to tag every false positive. 
However, events in which signal is observed associated with a pair of excited spectator qubits can be rejected as due to thermal error. Additionally, the rate of spectator pair excitations provides a data-driven measurement of the thermal false positive rate, enabling accurate background estimation independent of signal measurements. 

\section{Benchmarking Entangled Qubits against Cavity Experiments}
\label{sec:cavity_qubit_comparison}

As mentioned in the introduction, searches for wave-like dark matter which includes axions and dark photons in the 1 to tens of GHz are carried out in cavity-based experiments \cite{ADMX:2018gho,ADMX:2021nhd,ADMX:2025vom, Cervantes:2022gtv, dixit2021searching}. 
In these cavity haloscopes, the resonant conversion of axions to microwave photons is being searched under the influence of a strong magnetic field. Haloscopes can also search for dark photons converting to photons, which can occur independently of a magnetic field. Most haloscope efforts read out power emitted from the cavity classically, typically being limited by thermal noise. Even at very low temperature, the power will be limited by the $\hbar\omega/2$ that is vacuum fluctuations, the so-called standard quantum limit (SQL). A promising path to improve beyond the SQL is to use photon counting~\cite{Lamoreaux2013}. 
A dark photon search was recently demonstrated at a single frequency in a cavity-qubit system~\cite{dixit2021searching}. 
Advances in high coherence cavity-qubit systems~\cite{Milul2023, Kim:2025ywx} make this a promising direction for sensing.

One might ask how cavity experiments would compare to the use of qubits with the entangled protocols, particularly as superconducting qubits live in the same frequency range. We present a comparison here between the scan rates of the entanglement-based qubit protocol and cavity based sensing in a single cavity. On the cavity side we will consider both thermal power readout and photon counting. Because the qubit case is more comparable to photon counting in cavities, we will focus on this case in the text and defer the relevant formulae for the thermal case to \App{app:thermal_cavity}. As we are considering superconducting qubits, they are not the ideal platform for axion searches because of their limited operability in presence of a magnetic field. There has been some ongoing work towards realizing such a possibility \cite{Chen:2024aya}, however in this section we will limit our analysis to dark photons as a benchmark.

\subsection{Cavity scan rate}

The Hamiltonian of a cavity coupled to dark photon dark matter, when considering the transition of ground state $|0\rangle$ to a single photon Fock state $|1\rangle$, takes the same form as our qubit Hamiltonian Eq.~(\ref{eq:DM-qubit_Heff}), but with a coupling of 
\begin{equation}
    \eta_\mathrm{cav}= 
     \frac{1}{2} e \epsilon_{A'}\, \sqrt{\rho_\mathrm{DM}G \,V_\mathrm{cav}\, \omega} \cos \theta,
\end{equation} 
where $V_\mathrm{cav}$ is the cavity volume, and $G$ is a mode-dependent geometric factor of order 0.1-1~\cite{Cervantes:2022gtv} for a good mode choice. The probability to excite the cavity from the vacuum~$|0\rangle$ to a Fock state with a single photon~$|1\rangle$ is thus identical to the single qubit excitation probability Eq.~(\ref{eq:Psig}), with the substitution of $\eta\to\eta_\mathrm{cav}$.  The scan rate is also similar in form to the single qubit expression, 
\begin{equation}
\label{eq: cavity_scanrate}
\mathrm{SR}_\mathrm{cav}=
    \frac{\eta_{\rm cav}^{4} \tau_\mathrm{\small DM}^2}{(\snr)^{2} p^\mathrm{(cav)}_{\rm error}}.
\end{equation} 
where $p^\mathrm{(cav)}_{\rm error}$ captures the probability of a dark count.
We consider a cavity experiment using photon counting method to go beyond the thermal noise limits. A benchmark demonstration of such a method is shown in ref.~\cite{dixit2021searching}, where a transmon qubit is coupled dispersively with the cavity, which experiences a frequency shift depending on the number of photons in the cavity. 
This is a Quantum Non-demolition (QND) measurement of the number of photons in the cavity allowing for a reduced readout error rate. It is thus justified to assume that $p^\mathrm{(cav)}_{\rm error}$ is dominated by the low thermal error rate in the cavity-qubit system. In this demonstration a dark count rate of order 1~Hz was achieved, corresponding to $p^\mathrm{(cav)}_{\rm error}\sim 10^{-3}$. This corresponds to an improvement upon the SQL rate by a factor of 1000. In the figures below we will use this dark count rate as a benchmark value for cavities.

Photon counting in cavities is a fairly recent development for dark matter searches. To complete the picture we also quantify the scan rate of ``thermal'' cavity searches, where power is read out, following~\cite{Kim:2020kfo} in \App{app:thermal_cavity}. 

In the following subsection we will draw comparisons between benchmark cavity and qubit systems across a span of frequencies. For this purpose we will assume that the cavity volume scales with the resonant frequency as $V\propto \omega^{-3}$. This benchmark scaling can be varied, for example by using multi-cell geometries~\cite{Jeong:2017hqs, Jeong:2020cwz}.

\subsection{Comparison of cavities and qubits}

For simplicity, we begin by comparing the scan rates in the case where the DM coherence time is shorter than that of the GHZ state,  
$\tau_\mathrm{\small DM} \lesssim \tau_\mathrm{\small GHZ}=\tau_q/n_q$, and the integration time is taken to be of order $\tau_\mathrm{\small DM}$. 
Assuming small readout error, the entangled protocol scan rate in this regime shows an $n_q^3$ scaling, as shown in \Eq{eq:entangled_n3_scaling}. Since the photon counting cavity has a similar form, the ratio of entangled-qubit to cavity scan rate 
at the same target sensitivity to the gauge-kinetic mixing parameter $\epsilon$
has the form 
\begin{equation}
    \label{eq:SR-ratio}
\frac{\mathrm{SR}_\mathrm{qubit}}{\mathrm{SR}_\mathrm{cav}}= \left(\frac{\eta^4_\mathrm{qubit}}{\eta^4_\mathrm{cav}}\right)\left(
    \frac{p_\mathrm{error}^\mathrm{(cav)}}{p_\mathrm{error}^\mathrm{(qubit)}}\right)
    n_q^3 \approx 
    \left(\frac{V_\mathrm{eff}^2}{V_\mathrm{cav}^2}\right)\left(
    \frac{p_\mathrm{error}^\mathrm{(cav)}}{p_\mathrm{error}^\mathrm{(qubit)}}\right)  n_q^3 ,
\end{equation}
where we have assumed that the geometric factors of $\kappa$ and $G$, for qubits and cavities respectively, are of the same order. We have added the label `qubit' here to $\eta$ for clarity. We can discuss the three factors in \Eq{eq:SR-ratio} in turn, noting from the onset that there are interesting interplays among them:
\paragraph{Number of qubits:}
Qubit-based quantum computing systems range from approximately 100 qubits in current state-of-the-art devices to millions in the envisioned fault-tolerant era.
It is fortunate that superconducting qubits are compact devices that are on a path towards mass production at high yield. 
As we demonstrated above, adding more qubits is most valuable so long as the coherence time of the GHZ state is longer than $\tau_\mathrm{\small DM}$. 
As an example, for DM frequency of $\sim$ 6~GHz, $\tau_\mathrm{\small DM}\sim 20~\mu\mathrm{s}$\footnote{We note that the precise definition of the DM coherence time varies by convention. For example the analysis of~\cite{Cheong:2024ose} gives a factor of 4 or so longer $\tau_\mathrm{\small DM}$ than this estimate. This does not change the qualitative picture we present here.}, qubit $T_1$ and $T_2$ lifetimes of order 1~ms allow for up to 50 entangled qubits while naively maintaining $\tau_\mathrm{\small GHZ}\gtrsim\tau_\mathrm{\small DM}$. The factor of $n_q^3$, therefore, contributes to sizable search speedup. If gate errors are low, further gains can be had with an $n_q^2$ scaling as shown in \Fig{fig:scanrate_ratio} to the right of the red line.
As we will briefly touch upon in the discussion, multi-cavity systems and protocols yield further advantage on the cavity side, that are beyond our scope here.
\paragraph{Volume:}
Some of the gain from higher qubit number is offset by their smaller effective volume compared to cavities. 
For superconducting qubits, the effective volume $V_{\rm eff} = d^2 C$ is, roughly speaking, set by the volume enclosed within the capacitor plates of the qubit.
Converting capacitance $C\sim 0.1~\mathrm{pF}$ to natural units and assuming a generous $d\sim 300~\mu$m for modern qubits gives an effective volume $V_\mathrm{eff}\sim 1~\mathrm{mm}^3$ per qubit. It should be noted that custom designed qubits can be optimized to have larger effective volumes in principle.
By comparison, the 6~GHz cavity of~\cite{dixit2021searching} has a volume of $V_\mathrm{cav}\sim 10~\mathrm{cm}^3$, a factor of $10^4$ larger.
On the other hand, qubits can be tuned in a more passive and volume independent way~\cite{Koch:2007hay, Hutchings2017FluxIndependent}, by changing the flux for example. By comparison, tuning a cavity in a sizable way requires moving parts and changing the volume\footnote{More passive cavity tuning has been demonstrated but over a small frequency range~\cite{Zhao:2025thg}}. 
\paragraph{Errors:} 
The last factor in the ratio of scan rates is the probability for error. 
In \Eq{eq:SR-ratio}, 
$p_\mathrm{error}^\mathrm{(qubit)}$ is the effective per-qubit error probability, typically dominated by thermal and gate errors. 
The error rate may be effectively lowered by monitoring the spectator qubits as discussed in \Sec{sec:error}. The ultimate comparison of error rates will eventually depend on whether quantum error correction (QEC) that is suitable for a DM search can be implemented, both on the cavity and the qubit side. 
Of course, QEC incurs substantial overhead due to the large ratio of physical to logical qubits.
\\[2pt] 

Having discussed the various factors in abstraction, we present a more quantitative (though still rough) look at the state of the qubit-cavity comparison, as a benchmark.
\begin{figure}[hbt]
    \centering
    \subfloat[High qubit-coherence scenario]{\includegraphics[width=0.95 \linewidth]{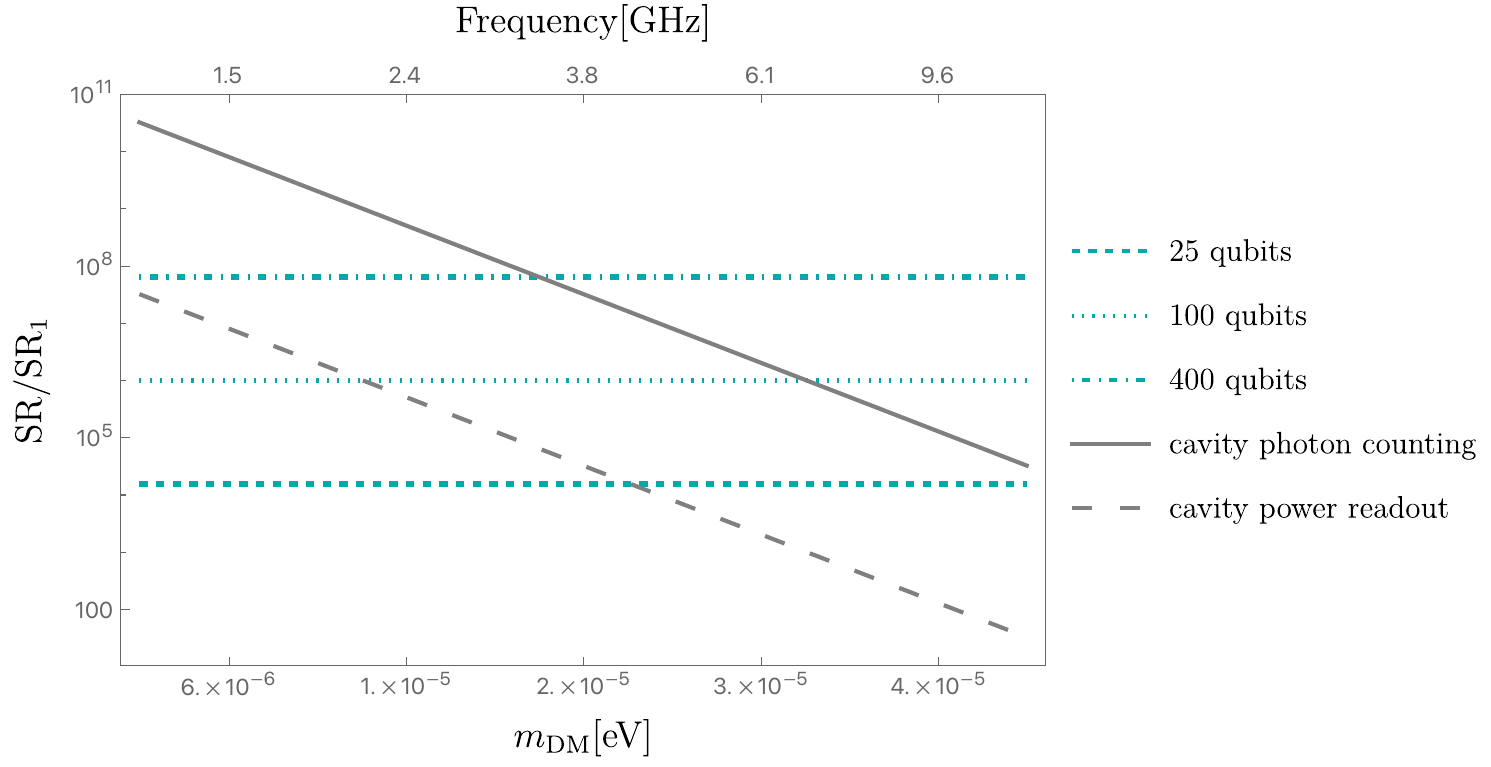}}
    \vspace{1.5em}
    \subfloat[State-of-the-art scenario]{\includegraphics[width=0.95 \linewidth]{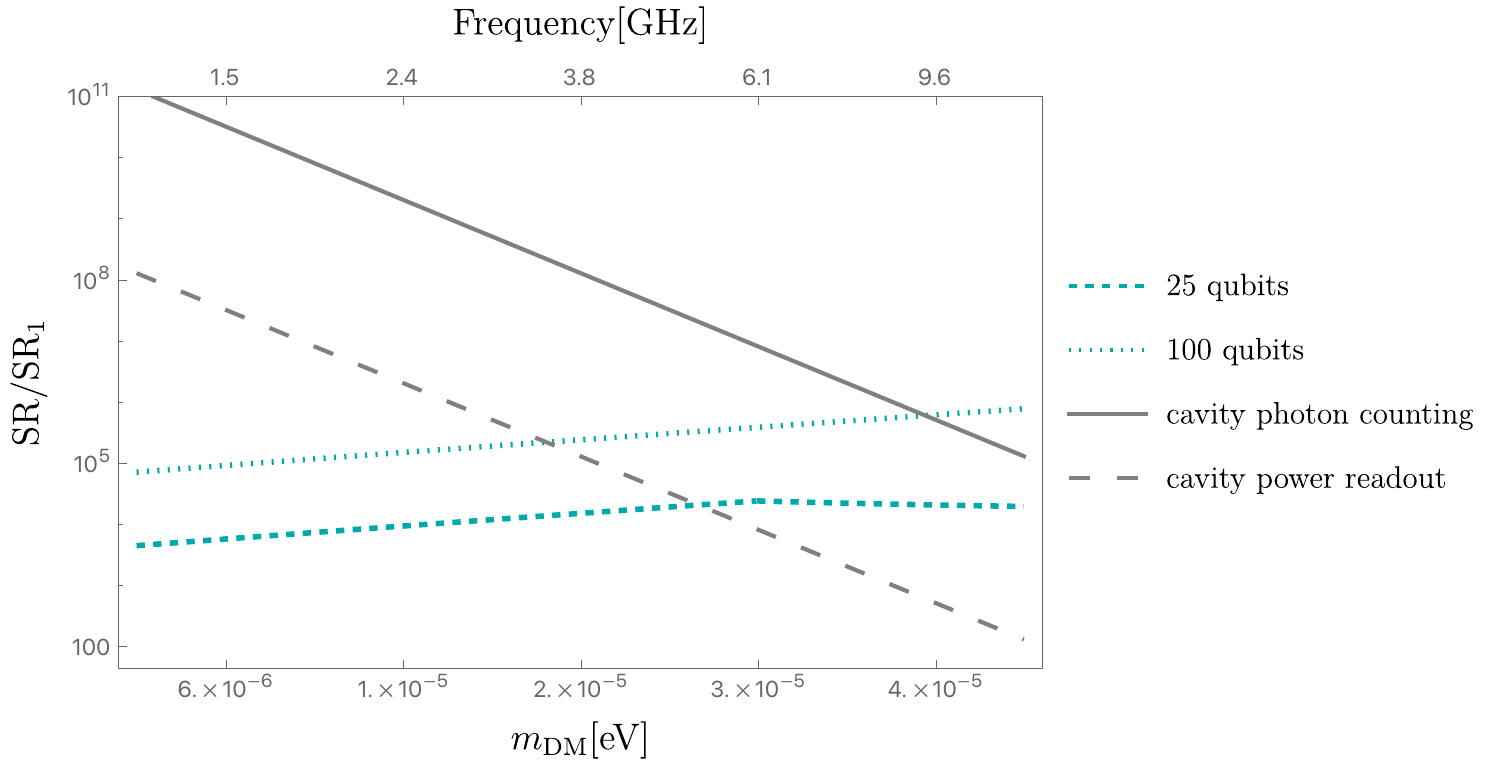}}
    \caption{Scan rates for cavities and the entangled-state protocol (normalized by the single-qubit scan rate) as a function of DM mass.
    The benchmark effective volume for the qubit $V_\mathrm{eff}=1$~mm$^3$ is taken to be independent of frequency, whereas the cavity volume is taken to scale as $\omega^{-3}$.
    We use ref.~\cite{dixit2021searching} as the benchmark for photon-counting cavity searches with the error rate of 0.1\%.
    For the thermal cavity, we assume an effective temperature of $10 ~ {\rm mK}$. In panel (a) we assume high qubit coherence and gate fidelity such that the exposure time is set by $\tau_\mathrm{\small DM}$ and the single-qubit error probability is below 0.25\%. In panel (b) we fix the qubit coherence $\tau_q \sim$ 1~ms, and set the exposure time to be $\min [ \tau_\mathrm{\small DM}, \tau_\mathrm{\small GHZ} ]$. 
    } \label{fig:scanrate_comparison}
\end{figure}
In \Fig{fig:scanrate_comparison} we plot scan rates of entangled qubits and cavities normalized to that of a single qubit as a function of $m_{\rm DM}$. 
We are interested in the $\sim 0.5 \times 10^{-5} ~{\rm eV}$ to $\sim 0.5 \times 10^{-4} ~{\rm eV}$ range as this corresponds to the current state-of-the-art ${\rm GHz}$-scale qubit frequencies. 
In the top panel, we consider an optimistic scenario with long coherence times and high gate fidelities, where the exposure time is set by $\tau_\mathrm{DM}$. 
We assume benchmark error probabilities of 0.1\% for photon-counting cavities and require single-qubit error probabilities below 0.25\%. 
For power readout cavities (detailed in Appendix~\ref{app:thermal_cavity}), we take a benchmark temperature of 10~mK, and assume that the experiment is SQL limited (which corresponds to an order one error probability coming from vacuum fluctuations). 
In the bottom panel we take current state-of-the-art qubit coherence times of $\tau_q\sim 1$~ms (see for example~\cite{somoroff2103millisecond, Bal:2023ccn, Bland:2025vvi}), picking an exposure time to be the shorter among the DM and GHZ coherence times. 
In this panel we take the probability of qubit gate error to be 1\%.  
The ability to scan qubits over a wide range passively can translate to an advantage over a cavity search under our benchmark assumptions.
We find that a GHZ protocol employing 100 qubits can be comparable or exceed the benchmark cavity search for DM masses of 30-40~$\mu$eV, depending on coherence and error rate assumptions. 

\begin{figure}[hbt!]
    \centering    
    \includegraphics[width=0.95 \linewidth]{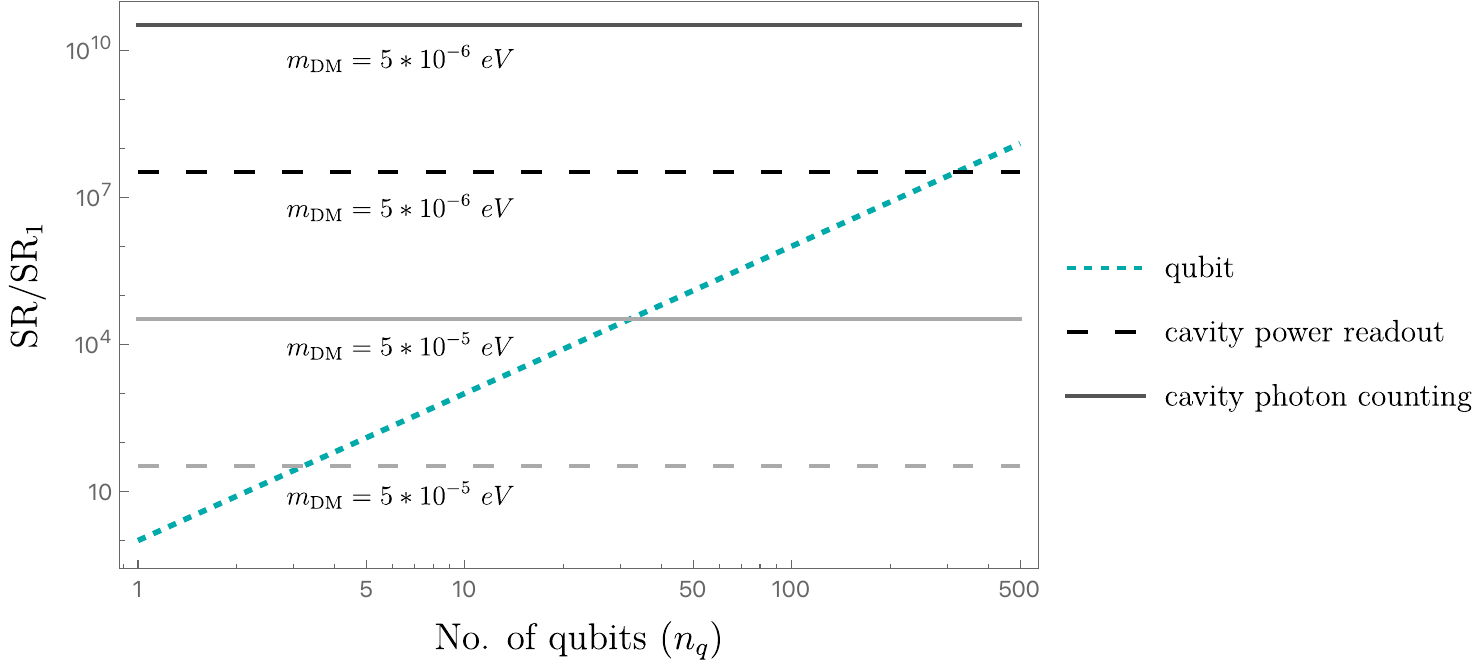}
    \caption{Scan rates for cavities and the entangled-state protocol (normalized by the single-qubit scan rate) as a function of number of qubits. Here we can see that the scan rate grows as $n_q^3$ as we are zooming in on the thermally dominated regime of Fig. \ref{fig:scanrate_ratio} for the entangled qubit case comparing to the benchmark cavity searches at specific target DM frequencies.  
    }
    \label{fig:scanrate_comparison_no._of qubits}
\end{figure}

In Fig. \ref{fig:scanrate_comparison_no._of qubits}, we show the normalized scan rate as a function of the number of qubits. We assume here that the qubit errors are thermal, scaling linearly with the number of qubits and exposure time, and that the GHZ coherence time is longer than $\tau_\mathrm{\small DM}$. In this regime the scan rate scales as $n_q^3$. At a dark matter mass of 50 $\mu$eV, 50 high quality qubits can exceed the scan rate of the assumed cavity search.

We reiterate that these comparisons are not meant to capture the ultimate performance of either cavity or qubit based searches. Rather, 
they serve as a benchmark to allow for informed choices between cavity and qubit searches as both technologies progress. In the following concluding section we discuss several factors that may lead to further quantum advantage in cavity and qubit technology.


\section{Summary and Future Directions}
\label{sec:conclusion}
Our analysis demonstrates that entangled qubit protocols can achieve genuine quantum advantages in DM searches, but within specific parameter regimes defined by experimental limitations.

The entangled protocol preserves the search bandwidth independent of qubit number, confirming superlinear scan rate scaling and avoiding the power-bandwidth tradeoff that limits power-readout and photon-counting schemes. This advantage stems from the  phase-based readout of the protocol, which extracts information without drawing power from the DM field. However, the sensitivity of the entangled state to errors constrains its practical advantage. Any single error within the entangled state can produce false positives, causing total error rates to scale with qubit number. As shown in Fig. \ref{fig:scanrate_ratio}, the  advantage in entangled protocol disappears once total error rates become substantial, limiting the maximum number of qubits that can be added to gain entanglement-based advantage.

While comparing to the more established haloscope designs, qubits face an inherent volume disadvantage, but this gap narrows at higher DM masses where cavity volumes scale unfavorably. 
As superconducting qubit technology advances to higher frequencies \cite{Anferov:2024ntf, Anferov:2024ljv}, the higher DM mass range becomes increasingly suitable for qubit-based detection.
For state-of-the-art superconducting qubits, our analysis indicates that approximately 100 entangled qubits could match power-readout cavity performance at $\sim 20\,\mu$eV and photon-counting cavity performance at $\sim 40\,\mu$eV, as shown in \Fig{fig:scanrate_comparison}(b). With fewer qubits at $n_q = 25$, these crossover masses are $\sim 30\,\mu$eV and $\sim 70\,\mu$eV, respectively.

Several extensions could enhance the practicality of the protocol. Various practical issues come about when all qubits are required to be at the same frequency. It would therefore be interesting to develop a broadband protocol with multi-frequency qubits that can retain some advantage from coherent addition.
Extending entanglement concepts to cavity arrays might alleviate volume advantages while preserving coherent scaling benefits. Additionally, investigating error-resilient entangled states or incorporating quantum error correction could push the optimal qubit number beyond current error-limited thresholds.
Identifying scenarios where error signatures differ sufficiently from dark matter signals could enable active error correction during the detection protocol.

While entangled quantum sensors remain experimentally challenging, they offer a theoretically viable path toward enhanced dark matter detection, particularly in the higher-mass regime where conventional approaches face scaling disadvantages.

\section*{Acknowledgments}

We are greatful to Nikita Blinov, Daniel Bowring, Shion Chen, Aaron Cou, Christina Gao, Anna Grasselino, Ryan Janish, Ryuichoro Kitano, Alex Millar, Takeo Moroi, Yao Lu, and Alex Romanenko for useful discussions.
This work was supported by a Department of Energy (DOE) QuantISED grant. 
The work of RH is also supported by the DOE Office of Science, National Quantum Information Science Research Centers, Superconducting Quantum Materials and Systems Center (SQMS) under contract number DE-AC02-07CH11359. The work of SG 
 was also supported
by the DOE, Office of Science, National Quantum Information Science Research Center,
Quantum Science Center (QSC) under contract number~DE-AC02-07CH11359. Fermilab is managed by FermiForward Discovery Group, LLC, acting under Contract No. 89243024CSC000002.
AB is also supported by DOE grant DE-SC0009924 at the University of Chicago.

\appendix

\section{Volume Scaling of Power}
\label{app:volume_scaling}

Here we show that the power deposited by DM in a passive detector (no amplification or injection of energy) scales with the volume of the detector. This result will also apply to photon counting.
In steady state, the Poynting theorem relates the rate of work done by the DM current to the rate of energy loss from the detector:
\begin{equation}
\label{eq:poynting}
    \int d^3 x \, \vec{J}_\mathrm{DM}\cdot \vec{E} = \frac{\omega^2}{Q} \int d^3x\, |\vec{E}|^2 + P_\mathrm{readout} .
\end{equation}
The left-hand side represents power delivered by the DM field, while the right-hand side consists of power dissipated in the detector (parameterized by quality factor $Q$) plus power extracted for readout.

The electric field in the signal mode can be decomposed into a time-dependent amplitude and a unit-normalized spatial profile:
\begin{equation}
    \vec{E} \equiv e(t)\hat{E}(\vec{x}),
\end{equation}
where the amplitude satisfies $|e|^2=\omega N_\mathrm{photons}$ (the energy stored for $N_\mathrm{photons}$ photons in the mode), and the spatial profile satisfies $\int |\hat{E}|^2 d^3x  = 1$. The mode profile $\hat{E}$ is a smooth function with average value of order $V^{-1/2}$.
Rewriting Equation~\ref{eq:poynting} gives:
\begin{equation}
    P_\mathrm{readout} = g_\mathrm{eff}\sqrt{\rho_\mathrm{DM} V\omega N_\mathrm{photons}} - \frac{\omega}{Q}N_\mathrm{photons}.
\end{equation}
The readout power has a maximum at finite photon number, beyond which cavity losses increase faster than the signal grows:
\begin{equation}
    P_{\rm readout} \leq \frac{g_{\rm eff}^2 \rho_{\rm DM}Q V}{4} .
\end{equation}
The main takeaway is that the optimal readout power scales linearly with total volume $V$. This establishes a fundamental constraint on power-based detection schemes, as well as on photon counting in sensors with pure Fock states.


\section{Calculation of the Excitation Probability Off Resonance}
\label{app:excitation_prob}

Here we derive the excitation probability for the entangled protocol at arbitrary detuning, extending the on-resonance result from \Sec{sec:entangled_protocol}.

As shown in \Sec{sec:dm_interaction}, the DM interaction with a qubit is described by a unitary in the computational basis:
\begin{align}\label{eq:U_DM}
    U_{\rm DM} &= \begin{pmatrix}
        e^{- it \Delta_\omega/2} (\cos \delta + i \sin \delta \sin \theta) & i e^{-i \varphi} e^{- it \Delta_\omega/2} \cos \theta \sin \delta \\
        i e^{i \varphi} e^{it \Delta_\omega/2} \cos \theta \sin \delta & e^{ it \Delta_\omega/2} (\cos \delta - i \sin \delta \sin \theta) ,
    \end{pmatrix}
\end{align}
where $\delta = \frac{\sqrt{\Delta_\omega^2+ 4 \eta^2}}{2} t$, $\theta = \tan^{-1} (\Delta_\omega/ 2 \eta)$, and $\Delta_\omega = \omega - m_{\rm DM}$ is the detuning.

For analyzing the entangled protocol, it is more convenient to express $U_{\rm DM}$ in terms of Pauli matrices:
\begin{align}
    U_{\rm DM} &= U_1 \mathbb{1} + U_x X +U_y Y+ U_z Z
\end{align}
where the coefficients are:
\begin{align}
    U_1 &= \cos\left(\frac{\Delta_\omega t}{2}\right) \cos\delta + \sin\left(\frac{\Delta_\omega t}{2}\right) \sin\delta \sin\theta, \notag \\ 
    U_x &= i \sin\delta \cos\theta \cos\left(\varphi+ \frac{\Delta_\omega t}{2}\right), \notag \\
    U_y &= i \sin\delta \cos\theta \sin\left(\varphi+ \frac{\Delta_\omega t}{2}\right), \notag \\
    U_z &= i \left(- \sin\left(\frac{\Delta_\omega t}{2}\right) \cos\delta+\cos\left(\frac{\Delta_\omega t}{2}\right) \sin\delta \sin\theta \right).
\end{align}

In the entangled protocol, we work in the $X$-basis states where $|\pm\rangle = (|0\rangle \pm |1\rangle)/\sqrt{2}$. The action of Pauli matrices on these states is: $X | \pm \rangle = \pm | \pm \rangle$, $Y | \pm \rangle = \mp i | \mp \rangle$, and $Z | \pm \rangle =  | \mp \rangle$. 
Therefore, the action of $U_{\rm DM}$ on the $X$-basis states can be written as:
\begin{align}
    U_{\rm DM} | \pm \rangle = C_{1\pm} | \pm \rangle + C_{2\pm} | \mp \rangle ,
\end{align}
where we have defined:
\begin{align}
    C_{1\pm} = U_1 \pm U_x, \qquad C_{2\pm} =  U_z \mp i U_y .
\end{align}

The GHZ-like state evolves under $U_{\rm DM}$ acting on each qubit. Since all qubits experience the same DM field, we have:
\begin{align}
    (U_{\rm DM} |+\rangle)^{\otimes n_q} &= (C_{1+} |+\rangle + C_{2+} |-\rangle )^{\otimes n_q} \nonumber \\
    &= \sum_{k=0}^{n_q} \left[C_{1+}^{k} C_{2+}^{n_q -k} \sum_{{\rm Config}(k)} |+\rangle^{\otimes k} |-\rangle^{\otimes (n_q -k)}\right],
\end{align}
where ${\rm Config}(k)$ denotes all configurations with $|+\rangle$ on $k$ sites and $|-\rangle$ on the remaining $(n_q-k)$ sites.
Similarly for $|-\rangle^{\otimes n_q}$, after a change of variables $k \rightarrow (n_q-k)$:
\begin{align}
    (U_{\rm DM} |-\rangle)^{\otimes n_q} = \sum_{k=0}^{n_q} \left[C_{2-}^{k} C_{1-}^{n_q -k} \sum_{{\rm Config}(k)} |+\rangle^{\otimes k} |-\rangle^{\otimes (n_q -k)}\right].
\end{align}
The full GHZ state after DM interaction is:
\begin{align}
    |\psi_{\rm DM} \rangle = \frac{1}{\sqrt{2}} \sum_{k=0}^{n_q} \left[(C_{1+}^{k} C_{2+}^{n_q -k} + C_{2-}^{k} C_{1-}^{n_q -k}) \sum_{{\rm Config}(k)}  |+\rangle^{\otimes k} |-\rangle^{\otimes (n_q -k)}\right].
\end{align}
After applying the serial CNOT gates, we obtain:
\begin{align}
    |\psi_{f}\rangle = \hat{O}_{\rm CNOTs}|\psi_{\rm DM}\rangle.
\end{align}
Calculation can be simplified by noticing that within $| \psi_f \rangle$, states in ${\rm Config}(k)$ and ${\rm Config}(n_q-k)$ can be paired.
To understand how states pair up, consider a general $n_q$-qubit eigenstate in the $X$-basis: $|x\rangle \equiv |x_1, x_2, \cdots, x_{n_q} \rangle$ where $x_i = \pm$. For convenience, we map $+ \leftrightarrow 0_{x}$ and $- \leftrightarrow 1_{x}$ (the subscript $x$ distinguishes this from the computational basis). The serial CNOTs act as:
\begin{align}\label{eq:CNOT_general_Xstate}
    |x\rangle \xrightarrow{\rm CNOTs} |x_1, (x_1 \oplus x_2), \cdots, (x_{n_q -1}\oplus x_{n_q}) \rangle ,
\end{align}
where $\oplus$ denotes addition modulo 2.
Now consider the complementary state $|\bar{x}\rangle = |(x_1 \oplus 1_x), (x_2 \oplus 1_x), \cdots, (x_{n_q} \oplus 1_x)\rangle$ (obtained by flipping all $+$ and $-$). Under serial CNOTs:
\begin{align}\label{eq:CNOT_complement_Xstate}
    |\bar{x}\rangle \xrightarrow{\rm CNOTs}& |(x_1 \oplus 1_x), (x_2 \oplus 1_x \oplus x_1 \oplus 1_x), \cdots, (x_{n_q -1} \oplus 1_x \oplus x_{n_q} \oplus 1_x)\rangle \nonumber \\
    =& |(x_1 \oplus 1_x), (x_1 \oplus x_2), \cdots, (x_{n_q -1} \oplus x_{n_q}) \rangle.
\end{align}

Comparing \Eqs{eq:CNOT_general_Xstate}{eq:CNOT_complement_Xstate}, we see that after CNOTs, the last $(n_q-1)$ qubits in $|x\rangle$ and $|\bar{x}\rangle$ are identical, while the first qubit is complementary. Therefore, for every state in ${\rm Config}(k)$, there exists a complementary state in ${\rm Config}(n_q-k)$ that has the same product state on the last $(n_q-1)$ qubits and complementary states on the first qubit. These states can be paired such that their contribution to $|\psi_f\rangle$ looks like
\begin{align}
   &\frac{1}{2} \left\lbrace \left[ \paren{C_{1+}^{k} C_{2+}^{n_q -k}+ C_{2-}^{k} C_{1-}^{n_q -k}} + \paren{C_{1+}^{n_q -k} C_{2+}^{k}+ C_{2-}^{n_q -k} C_{1-}^{k}} \right] |0\rangle_{1} \right. \nonumber \\
   &\left. \pm \left[ \paren{C_{1+}^{k} C_{2+}^{n_q -k}+ C_{2-}^{k} C_{1-}^{n_q -k}} - \paren{C_{1+}^{n_q -k} C_{2+}^{k}+ C_{2-}^{n_q -k} C_{1-}^{k}} \right] |1\rangle_{1} \right\rbrace \otimes |\text{spectators}\rangle ,
\end{align}
where the $\pm$ sign depends on whether the first qubit in the Config($k$) state is $|+\rangle_1$ or $|-\rangle_1$.

Tracing over the $(n_q-1)$ spectator qubits and measuring the first qubit gives the general excitation probability:
\begin{align}\label{eq:Pexcite_exact}
    p_{\rm sig} = \frac{1}{8} \sum_{k=0}^{n_q} \binom{n_q}{k} \left| \left(C_{1+}^{k} C_{2+}^{n_q -k}+C_{2-}^{k} C_{1-}^{n_q -k}\right) - \left(C_{1+}^{n_q -k} C_{2+}^{k}+C_{2-}^{n_q -k} C_{1-}^{k}\right) \right|^2 .
\end{align}
This exact expression accounts for arbitrary detuning and DM phase $\varphi$. We now examine two limiting cases.

\textbf{On-resonance limit ($\Delta_\omega = 0$, $\delta \ll 1$):} On resonance with small phase accumulation, the coefficients simplify to:
\begin{align}\label{eq:C12_onResonance}
    C_{1\pm} &= 1 \pm i \delta \cos\varphi -\delta^2 /2+ {\cal O}(\delta^3), \\
    C_{2\pm} &= \pm \delta \sin\varphi + {\cal O}(\delta^3)
\end{align}
Terms contributing to $p_{\rm sig}$ at ${\cal O}(\delta^2)$
only come from $k=0, 1, (n_q-1), n_q$. The contributions from $k=0$ and $k=n_q$ are equal, as are those from $k=1$ and $k=n_q-1$. This gives:
\begin{align}\label{eq:Pexcite_onResonance}
    p_{\rm sig}\big|_{\Delta_\omega =0} &=  \frac{1}{4} \left[ |C_{1-}^{n_q} - C_{1+}^{n_q}|^2 + n_q |C_{2-} C_{1-}^{n_q -1} -  C_{1+}^{n_q -1} C_{2+}|^2 \right] \nonumber \\   
    &= n_q^2 \delta^2 \cos^{2}\varphi + n_q \delta^2 \sin^{2}\varphi + {\cal O}(\delta^3).
\end{align}
The dominant term scales as $n_q^2$, confirming the quadratic enhancement from entanglement.

\textbf{Far off-resonance limit ($\Delta_\omega \gg \eta$):} For large detuning, we expand in the small parameter $\eta/\Delta_\omega$. To leading order:
\begin{align}\label{eq:C12_offResonance}
    C_{1\pm} &= 1 \pm i \frac{2\eta}{\Delta_\omega} \sin\left(\frac{\Delta_\omega t}{2}\right) \cos\left(\varphi + \frac{\Delta_\omega t}{2}\right) + 
    \frac{1}{2}\left(\frac{2\eta}{\Delta_\omega}\right)^2 \sin\left(\frac{\Delta_\omega t}{2}\right)^2 +\cdots, \\
    C_{2\pm} &= \pm \frac{2\eta}{\Delta_\omega} \sin\left(\frac{\Delta_\omega t}{2}\right) \sin\left(\varphi + \frac{\Delta_\omega t}{2}\right) + i \left(\frac{\eta^2}{\Delta_\omega^2}\right)\paren{\Delta_\omega t-\sin(\Delta_\omega t)} + \cdots.
\end{align}
For leading order terms that dominantly contribute to signal, the above equations have the same structure as the on-resonance case with the substitutions $\delta \rightarrow (2\eta/\Delta_\omega) \sin(\Delta_\omega t/2)$ and $\varphi \rightarrow \varphi + \Delta_\omega t/2$. Therefore:
\begin{align}\label{eq:Pexcite_Detuned}
    p_{\rm sig}\big|_{\Delta_\omega \gg \eta} = \frac{4 \eta^2}{\Delta_\omega^2} \sin^{2} \left(\frac{\Delta_\omega t}{2}\right)  \left[n_q^2 \cos^{2}\left(\varphi+ \frac{\Delta_\omega t}{2}\right)  + n_q \sin^{2}\left(\varphi+ \frac{\Delta_\omega t}{2}\right)\right]+ {\cal O}\paren{\frac{\eta^3}{\detun^3}}.
\end{align}

The key observation is that $p_{\rm sig}$ drops when $\Delta_\omega \gtrsim t^{-1}$, independent of $n_q$. In the intermediate regime $\eta < \Delta_\omega < t^{-1}$, the signal probability $p_{\rm sig} \propto n_q^2 (\eta t)^2 \cos^{2}\varphi$ matches the on-resonance result. However, for $\Delta_\omega \gtrsim t^{-1}$, the oscillating $\sin^2(\Delta_\omega t/2)$ factor suppresses the signal.
This establishes that the bandwidth is determined solely by the exposure time $t$ and is independent of the number of qubits $n_q$. The entangled protocol preserves the bandwidth $\Delta\omega \sim 1/t$ while achieving the $n_q^2$ signal enhancement, confirming that it avoids the power-bandwidth trade-off discussed in \Sec{sec:bandwidth}.

\section{Effect of Thermal Errors in the Entangled-State Protocol}
\label{app:error}

Here we analyze the effect of thermal relaxation errors on the entangled-state protocol.
Thermal excitation errors can be treated analogously and lead to similar conclusions.
Consider a thermal relaxation error occurring on the $j^{\rm th}$ qubit at time $t_1$ during the protocol. 
We model qubit de-excitation as emission of a photon that entangles the qubit with the environment:
\begin{align}\label{eq:thermal_error_model}
    &\ket{0}_j \, \ket{0}_\gamma \rightarrow  \ket{0}_j 
    \, \ket{0}_\gamma ,\quad
    \ket{1}_j \, \ket{0}_\gamma \rightarrow  \ket{0}_j 
    \, \ket{1}_\gamma ,
\end{align}
where $|0\rangle_{\gamma}$ and $|1\rangle_{\gamma}$ denote the vacuum and single-photon states of the environment.
At time $t_1$ just before the error occurs, the state is:
\begin{align}
    \ket{\psi(t_1)} &= \frac{1}{\sqrt{2}} \left[ e^{i n \delta_1} \ket{+}^{\otimes (j-1)} \paren{\frac{\ket{0}_j + \ket{1}_j}{\sqrt{2}}} \ket{+}^{\otimes (n-j)}\right. \nonumber \\
    & \hspace{3.5em} \left.
    + e^{-i n \delta_1} \ket{-}^{\otimes (j-1)} \paren{\frac{\ket{0}_j - \ket{1}_j}{\sqrt{2}}} \ket{-}^{\otimes (n-j)} \right] \otimes \ket{0}_\gamma \, ,
\end{align}
where $\delta_1 \approx \eta t_1$ is the phase accumulated up to time $t_1$ and we have expanded $\ket{\pm}_j$ in the computational basis.
Applying the error model from \Eq{eq:thermal_error_model}, the state after de-excitation becomes:
\begin{align}
    \ket{\psi(t_1)}_{\rm error} &= \frac{1}{2} \left[ e^{i n \delta_1} \ket{+}^{\otimes (j-1)} \lbrace\ket{0}_j \otimes (\ket{0}_\gamma+\ket{1}_\gamma)\rbrace \ket{+}^{\otimes (n-j)} \right. \nonumber \\
    & \hspace{2.5em} \left.+ e^{-i n \delta_1} \ket{-}^{\otimes (j-1)} \lbrace\ket{0}_j \otimes (\ket{0}_\gamma-\ket{1}_\gamma)\rbrace\ket{-}^{\otimes (n-j)} \right] \nonumber \\
    &=\frac{1}{2} \left[ e^{i n \delta_1} \ket{+}^{\otimes (j-1)} \lbrace (\ket{+}_j +\ket{-}_j )\otimes \ket{+}_\gamma\rbrace \ket{+}^{\otimes (n-j)}  \right. \nonumber \\
    & \hspace{2.5em} \left.+ e^{-i n \delta_1} \ket{-}^{\otimes (j-1)} \lbrace(\ket{+}_j +\ket{-}_j ) \otimes \ket{-}_\gamma\rbrace\ket{-}^{\otimes (n-j)} \right] ,
\end{align}
where we have defined $\ket{\pm}_\gamma = (\ket{0}_\gamma \pm \ket{1}_\gamma)/\sqrt{2}$.
We continue the protocol for the rest of the time, $t_2=t_{\rm DM}-t_1$. 
During this period, each qubit accumulates additional phase  $\delta_2 = \eta t_2$. The state after full DM exposure is:
\begin{align}
    \ket{\psi_{\rm DM}} &= \frac{1}{2} \left[ e^{i n_q \delta} \ket{+}^{\otimes n_q}  \ket{+}_\gamma 
    + e^{i (n_q \delta-2\delta_2)} \ket{+}^{\otimes (n_q-1)} \ket{-}_j  \ket{+}_\gamma  \right. \nonumber \\
    & \hspace{2.5em} \left.
    + e^{-i (n_q \delta-2\delta_2)} \ket{-}^{\otimes (n_q-1)} \ket{+}_j  \ket{-}_\gamma 
    + e^{-i n_q \delta} \ket{-}^{\otimes n_q}  \ket{-}_\gamma \right],
\end{align}
where $\delta = \delta_1 + \delta_2$ is the full phase.
Next, we apply the serial CNOT gates to transfer phase information to the first qubit, followed by Hadamard gates on the spectator qubits to return them to the computational basis.
The final state becomes 
\begin{align}\label{eq:spectator_error}
    \ket{\psi_f} 
    &= \frac{1}{2} \left[ (\cos n_q \delta \,\lbrace \ket{0}_1 \ket{0}_\gamma + \ket{1}_1 \ket{1}_\gamma \rbrace + i \sin n_q \delta \,\lbrace \ket{1}_1 \ket{0}_\gamma + \ket{0}_1 \ket{1}_\gamma \rbrace  ) \otimes \ket{0}^{\otimes (n_q-1)}  \right. \nonumber \\
    & \hspace{2.5em} \left.
    + (\cos (n_q \delta-2 \delta_2)\, \lbrace  \ket{0}_1 \ket{0}_\gamma + \ket{1}_1 \ket{1}_\gamma \rbrace + i \sin (n_q \delta-2 \delta_2) \, \lbrace \ket{1}_1 \ket{0}_\gamma + \ket{0}_1 \ket{1}_\gamma \rbrace) \right. \nonumber \\
    & \hspace{3.5em} \left.
     \otimes \ket{0}^{\otimes (n_q-3)} \ket{1}_j \ket{1}_{j+1}
    \right] .
\end{align}
Since the emitted photon escapes to the environment and cannot be measured, we must trace over the photon degree of freedom.
Then the probability of seeing an excitation on the first qubit together with no excitation on the spectators is $1/2$. 
Excitation on consecutive qubits $j$ and $j+1$ is seen with probability $1/2$ and reveals the occurrence of error.
However, the probability of seeing the signal qubit excited is still $1/2$.
In both cases, the probability is independent of 
$\delta$, which demonstrates that thermal relaxation errors completely erase the DM signal, even if the error is detected via excitations on spectator qubits.
This contributes to the enhanced false positive rate discussed in \Sec{sec:noise_and_scanrate}. 
The errors are therefore particularly detrimental to the entangled protocol.
The probabilities derived here correspond to the entries in Table~\ref{tab:errors} for thermal relaxation errors.


\section{Power Readout from Cavities}
\label{app:thermal_cavity}

In this appendix we present the scan rate for the DM searches which employ the readout of power, in the thermal regime, which is the most common search strategy~\cite{Sikivie:1983ip}. 
The signal power 
in a dark photon haloscope is 
\begin{equation}
\label{eq: dark_photon_power}
P_{A'} = \epsilon_{A^\prime}^2 m_{A'} \rho_{A'}V_{\rm cav} \,\mathrm{min}(Q_{A'}, Q_\mathrm{cav})\, C^{(A')}_\mathrm{mode}
\end{equation}
where $\epsilon_{A^\prime}$ is the small gauge-kinetic mixing, while $\rho_{A'}$ and $m_{A'}$ are the dark photon DM energy density and mass, respectively. The absence of a magnetic field allows one to use superconducting ultra-high $Q$ cavities, benefiting the search~\cite{Cervantes:2022gtv}. According to \cite{Kim:2020kfo}, a more accurate description of the the quality factor term would be  $\frac{Q_l Q_{\rm A^\prime}}{Q_l +Q_{\rm A^\prime}}$, which we will adapt here for our analysis. A general form for the scan rate of these haloscopes where the power is readout has been presented in ref.~\cite{Kim:2020kfo}:
\begin{equation}
    \label{eq: cavity_scanrate_from_reference}
   \rm{SR_{\rm cav, power}} =\left.\frac{\Delta \omega}{\Delta t}\right|_{\rm cav, power} = \frac{1}{({\rm SNR})^2} \left(\frac{\epsilon_{A^\prime}^2 m_{A'} \rho_{A'}V }{k_B T_{\rm eff}}\right)^2 \left(\frac{\frac{\beta}{1+\beta}}{\frac{4 \beta}{(1+\beta)^2}+\lambda}\right)^2 \frac{Q_l Q_{A^\prime}^2}{Q_l +Q_{A^\prime}}.
\end{equation}
This scan rate expression can also be applicable to axions as shown in \cite {Kim:2020kfo}. $\lambda$ here characterizes the ratio between the added noise in the system in comparison to the thermal noise, $\lambda= T_{\rm add}/ T_{\rm eff}$. 
Here $T_{\rm eff}$ is the sum of the system temperature and another term accounting for the zero-point
fluctuations as illustrated below, which becomes important at frequencies where the thermal noise is sub-dominant,
\begin{equation}
    T_{\rm eff} \sim T_{\rm sys} \frac{\hbar \omega}{K_B T_{\rm sys}}\left(\frac{1}{e^{\frac{\hbar \omega}{K_B T_{\rm sys}}}-1}+\frac{1}{2}\right).
\end{equation}
For example, if we consider operating around 10 mK which is roughly equivalent to $10^{-6}  \rm{eV} $, at masses lower than that, $\hbar \omega< K_B T_{\rm sys}$ and $T_{\rm eff} \sim T_{\rm sys}$, whereas above that $\hbar \omega> K_B T_{\rm sys}$, leading to $T_{\rm eff} \sim \frac{1}{2} \frac{\hbar \omega}{K_B }$, which makes the noise term frequency, or mass, dependent.  Also, $\beta$ is the coupling strength to the receiver and $Q_l$ is the loaded quality factor related to the cavity quality factor $Q_{\rm cav}$ by $Q_l = Q_{\rm cav}/ (1+\beta)$. These parameters contribute to the G factor of the cavity as in \Sec{sec:cavity_qubit_comparison}.  As mentioned, this is primarily characterized as the scan rate of a cavity experiment where the limiting noise source is the thermal noise. In \Sec{sec:cavity_qubit_comparison} we have emphasized the other scenario with photon-counting readout for cavities, where the limiting factor is not the thermal noise but it is rather sub-dominant to other sources of noise such as readout error.
The scan rate of the photon counting cavities laid out in \Eq{eq: cavity_scanrate}, which resembles the qubits scan rate, captures the similar frequency dependence and scaling presented here in \Eq{eq: cavity_scanrate_from_reference}, where the $p_\mathrm{error}^\mathrm{(cav)}$ in \Eq{eq: cavity_scanrate} represents the improvement factor over SQL complementing the $T_{\rm eff}$ description in Eq. (\ref{eq: cavity_scanrate_from_reference}).

\bibliographystyle{JHEP}
\bibliography{refs.bib}

\end{document}